\newif\ifonecol
\onecoltrue

\ifonecol
\documentclass[12pt,draftclsnofoot,onecolumn]{IEEEtran}
\else
    \documentclass[journal,comsoc]{IEEEtran}
\fi
\usepackage{amsthm} 

\usepackage{amsmath}
\usepackage[cmintegrals]{newtxmath}
\usepackage{bm}
\usepackage{algorithmic}
\usepackage{array}
\usepackage{algorithm}
\usepackage{color}
\usepackage{makecell}
\usepackage{tabularx} 
\usepackage{cite}
\usepackage{multirow}
\usepackage{enumerate}
\usepackage{epsfig}
\usepackage[tight]{subfigure}

\renewcommand{\thefigure}{\arabic{figure}}

\begin{document}
\ifonecol
    \title{On the Capacity of MISO Channels \\with One-Bit ADCs and DACs }
    \vspace{-2mm}
\fi

\author{Yunseo Nam, Heedong Do, Yo-Seb Jeon, and Namyoon Lee
	\thanks{Y. Nam, H. Do, Y.-S. Jeon, and N. Lee are with the Department of Electrical Engineering, POSTECH, Pohang, Gyeongbuk 37673, South Korea  (e-mail: \{edwin624, doheedong, yoseb.jeon, nylee\}@postech.ac.kr).}
}
\vspace{-2mm}

\maketitle

\setlength\arraycolsep{2pt}
\newcommand{\argmax}{\operatornamewithlimits{argmax}}
\newcommand{\argmin}{\operatornamewithlimits{argmin}}
\makeatletter
\newcommand{\vast}{\bBigg@{3.5}}
\newcommand{\Vast}{\bBigg@{4.5}}
\makeatother
\vspace{-12mm}

\begin{abstract} 
\vspace{-0.2cm}
A one-bit wireless transceiver is a promising communication architecture that not only can facilitate the design of mmWave communication systems but also can extremely diminish power consumption. The non-linear distortion effects by one-bit quantization at the transceiver, however, change the fundamental limits of communication rates. In this paper, the capacity of a multiple-input single-output (MISO) fading channel with one-bit transceiver is characterized in a closed form when perfect channel state information (CSI) is available at both a transmitter and a receiver. One major finding is that the capacity-achieving transmission strategy is to uniformly use four multi-dimensional constellation points. The four multi-dimensional constellation points are optimally chosen as a function of the channel and the signal-to-noise ratio (SNR) among the channel input set constructed by a spatial lattice modulation method. As a byproduct, it is shown that a few-bit CSI feedback suffices to achieve the capacity. For the case when CSI is not perfectly known to the receiver, practical channel training and CSI feedback methods are presented, which effectively exploit the derived capacity-achieving transmission strategy.
\end{abstract}
\begin{IEEEkeywords}
\vspace{-0.3cm}
MISO channel, one-bit quantization, channel capacity, and spatial lattice modulation.\end{IEEEkeywords}
\vspace{-0.3cm}
\section{Introduction}
\subsection{Motivation}
The use of large bandwidths at mmWave bands is a key technology for future wireless systems supporting high data throughput \cite{Swindlehurst2014,Sun:12}.  Although the large signal bandwidth is able to provide a significant improvement of data rates, it complicates the design of analog and analog-digital mixed hardware components in transceiver. In particular, very high-speed digital-to-analog converters (DACs) and analog-to-digital converters (ADCs) at the transceiver can lead to significant power consumption \cite{Murmann,Walden:99, Singh:09}. For example, it has shown in \cite{Murmann} that an ADC consumes two Watts when quantizing a received signal with the sampling rate of 4 Gsamples per second and 12-bit precision per sample. In addition, multiple transceiver chains each with a large amount of antenna elements, i.e., hybrid-precoding \cite{Omar2014,Amed2014}, are essentially needed to compensate a high path loss existing in mmWave channels and to offer high data rates by simultaneously sending multiple data streams. As a result, the use of multiple transceiver chains in mmWave wireless systems additionally increase the circuit power consumption and the hardware cost.  

A simple solution for reducing the huge power consumption and the expensive hardware cost is to exploit low-precision DACs and ADCs at the transceiver. By reducing the number of quantization bits per sample, the power consumption can be decreased exponentially \cite{Hoyos_TWC2005,Blazquez2005,   Madhow2009}. Such low-power and cost-effective solution, however, significantly alters the characteristics of the wireless system due to the nonlinear distortion effects of transmit and receive signals; thereby, it changes the fundamental limits of communication rates and the practical communication schemes achieving these limits.

\vspace{-0.1cm}
\subsection{Related Works}
There have been an increasing research interest in both understanding the information theoretic limits and designing the practical communication schemes using low-resolution DACs or ADCs. Considering one extreme case in which a transmitter employs infinite-precision DACs, while a receiver uses one-bit ADCs, information theoretic limits were analyzed in \cite{Madhow2009,Mezghani2008,Mezghani2009,Nossek2006,Nossek20061,Mezghani2007,Mo2015_TSP,Mezghani2017}. For example, it was shown in \cite{Madhow2009} that an uniformly distributed quadrature phase shift keying (QPSK) signaling achieves the capacity of a quantized single-input single-output (SISO) additive white Gaussian noise (AWGN) channel. For a non-coherent communication system, i.e., no CSIR is available, a closed form expression of the capacity was derived as a function of the coherence time and the SNR by solving a convex optimization problem \cite{Mezghani2008}.  With the derived expression, it was demonstrated that the capacity-achieving input for a SISO Rayleigh-fading channel is the on-off QPSK.  For the coherent communication scenario when both CSIT and CSIR are available, the capacity expression of a MISO fading channel was derived in a closed form \cite{Mo2015_TSP}. Specifically, the capacity-achieving transmission method is to use the uniform QPSK modulation with the maximum ratio transmit (MRT) precoding. These results revealed that the use of a finite number of constellation points with the uniform distribution is optimal when the receiver has one-bit resolution. Such capacity-achieving input distribution is different from the Gaussian input distribution, which is known to be optimal when infinite-precision ADCs are employed.

 For the multiple-input multiple-output (MIMO) channel case, an exact channel capacity expression in a closed form is still unknown \cite{Nossek2006,Nossek20061,Mezghani2007,Mo2015_TSP, Mezghani2017, Mezghani2009}.  Instead of characterizing the exact channel capacity expression in a closed form, several works focused on the asymptotic characterization of the channel capacity of a MIMO channel \cite{Mo2015_TSP,Mezghani2007,Mezghani2009,Mezghani2017}. In \cite{Mo2015_TSP}, an upper and a lower bound of the capacity were characterized by using the hyperplane-cutting argument in the extremely high SNR regime. Furthermore, an asymptotical channel capacity expression was derived in the extremely low SNR regime by using a second-order expansion of the mutual information between the input and the output of the MIMO channel \cite{Mezghani2007,Mezghani2009,Mezghani2017}. One important finding was that the capacity loss caused by the one-bit quantization is unexpectedly small (1.96 dB) compared to the case of using infinite-precision ADCs in the low SNR regime. The common limitation of the aforementioned studies, however, is that they ignore the effect of low-precision DACs of the transmit chains.

Unlike the case of using one-bit ADCs at receivers only, a few works have focused on the capacity analysis for the case in which both the receiver and the transmitter are equipped with one-bit ADCs and DACs, respectively. In \cite{Gao2017,Gao2018}, the channel capacity of a real-valued MIMO channel have been characterized under the assumption of perfect CSIR only. Specifically, under the constraint of the binary phase shift keying (BPSK) constellation per each transmit antenna as a channel input, an asymptotical expression of the constraint-capacity is derived as a function of the ratio between the number of transmit and receive antennas and the SNR when the number of antennas goes to infinity. The key technique to find this capacity expression is the replica method \cite{Replica}, which has been widely used as a tool for statistical mechanics and information theory. Although the results in \cite{Gao2017,Gao2018} provide a useful guidance on how the capacity behaves in an asymptotical regime as a function of the important system parameters, the benefits of exploiting CSIT are not revealed. In addition, the major limitation of the works in \cite{Gao2017,Gao2018} is the assumption of BPSK signaling per transmit antenna as channel inputs. This assumption excludes the possibility of using spatial modulation methods \cite{Mesleh2008,Renzo2011,Mesleh2015,Yang2015,Ibrahim2016,Choi2018}, where information bits are jointly mapped across a set of activated indices of spatial, in-phase, and quadrature dimensions and the BPSK symbols of the activated dimensions. Under the one-bit DACs constraint, a set of all possible channel inputs is generated using spatial lattice modulation (SLM) method \cite{Choi2018}, which is known as a generalization of spatial modulation techniques \cite{Mesleh2008,Renzo2011,Mesleh2015,Yang2015,Ibrahim2016}. Since the joint information mapping by SLM maximizes the number of possible input alphabets, in general, it might achieve higher transmission rates compared to the case when using the BPSK signaling per transmit antennas as in \cite{Gao2017,Gao2018}.

  \begin{figure*}
    \centering
    \includegraphics[width=5.0in]{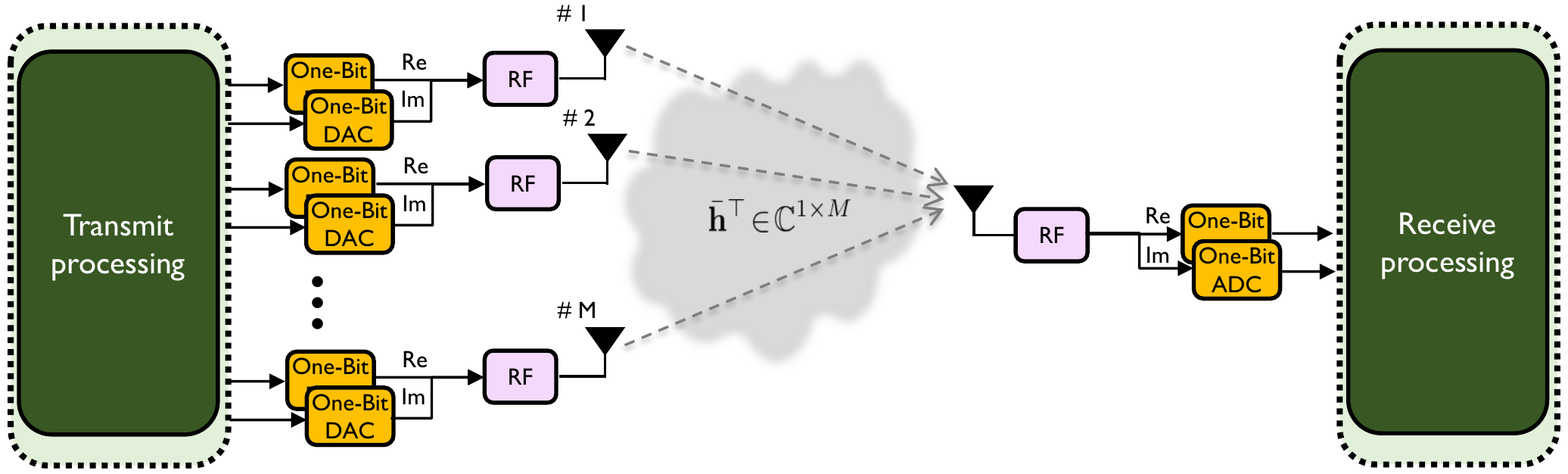}
    \vspace{-0.3cm}
    \caption{An illustration of the MISO system with one-bit transceiver.}
    \label{Fig1}
    \vspace{-0.6cm}
\end{figure*}
\vspace{-0.1cm}
  \subsection{Contributions} 
  In this paper, we consider a MISO channel in which a transmitter sends a message using $M$ multiple transmit antennas (or interchangeably transmit chains) each with one-bit DACs, and a receiver decodes the message with a single receive antenna with one-bit ADCs. Our goal in this paper is to understand the joint impact of one-bit ADCs and DACs on the capacity of a MISO channel when the perfect knowledge of CSIT and CSIR is available. In particular, we characterize the capacity expression with a closed form for the MISO channel with one-bit ADCs and DACs. The key idea to find the capacity expression is to construct the channel input vectors in an $2M$-dimensional input space $\{-1,0,1\}^{2M}$ by the joint information mapping across the transmit antennas, the real and the imaginary components \cite{Choi2018}. Then, we optimize the input distribution to maximize the mutual information between the channel inputs and outputs.

  We first start with the SISO channel with one-bit ADCs and DACs, i.e., $M=1$. One major finding is that the capacity-achieving transmission method is to adaptively exploit both QPSK and spatial modulation as a function of the channel amplitude, phase, and the SNR. This result is different from the capacity result of the SISO channel when using one-bit ADCs but employing infinite-precision DACs. In this case, the uniform QPSK constellation with precoding to align the channel phase is shown to be optimal. Another important observation is that one-bit CSIT is sufficient to achieve the capacity. This is quite appealing in practice, because the receiver needs to send back just binary information to the transmitter. Whereas, for the case of the SISO channel with one-bit ADCs and infinite-precision DACs, infinite amount of feedback bits are required to perform precoding for the phase alignment. By comparing the known capacity results, we provide a complete picture on how both one-bit ADCs and DACs changes the capacity of the SISO channel.

  We extend the results of the SISO case into the MISO case. For the MISO channel, it is possible to construct the channel input set with $9^M-1$ different non-zero input vectors by the joint information mapping rule in \cite{Choi2018}. Using this channel input set, we characterize the channel capacity by deriving the capacity-achieving input distribution under the average transmit power constraint. Our major finding is that the capacity-achieving transmission strategy is to send information bits by uniformly using four signal points in $\{-1,0,1\}^{2M}$. The optimal four signal points are selected among the SLM signals as a function of the channel and the SNR. In addition, it is shown that $\log_2\left(\frac{9^M-1}{4}\right)$ feedback bits are sufficient to achieve the capacity. This result is also different from the capacity result of the MISO channel with one-bit ADCs and infinite-precision DACs, in which the uniform QAM signaling with MRT beamforming is shown to be optimal.

  We also present a practical channel-training and CSI feedback method for the MISO channel with one-bit ADCs and DACs when perfect CSI is not available. The proposed idea is to exploit a supervised-learning approach in \cite{Jeon2017_SL,Jeon2018_SL,Jeon_So_Lee_2018}. In the phase of the channel-training, the transmitter sends the channel input vectors repeatedly, which are chosen from $\frac{9^M-1}{4}$ distinct subsets of SLM signal set. Using the received signals during the channel-training, the receiver empirically estimates the entropy functions for each subset. From the estimated entropy functions, the receiver selects the best subset that maximizes the channel capacity. Then, the index of the selected subset is sent back to the transmitter via a finite-rate feedback link. To reduce the channel-training and CSI feedback overheads, a sub-optimal channel-training and CSI feedback method is also proposed, in which only channel input vectors with the maximum instantaneous power level are sent during the channel-training phase. By simulations, it is shown that the proposed method is able to achieve the capacity within 0.2 bits/sec/Hz for all SNRs when $M=4$ and the total training-length of 320 and 4-bit CSI feedback are employed, respectively.

    Our paper is organized as follows. Section II describes the system model. In Section III, the capacity expression of the SISO channel with one-bit ADCs and DACs is characterized. Then, Section IV provides a closed form expression of the channel capacity when the transmitter has multiple antennas. When perfect CSI is not available, a practical channel-training and a CSI feedback method are presented in Section V. Some simulation results are provided to validate our results in Section VI. Section VII concludes the paper with some discussion about possible extensions.
    
\vspace{-0.1cm}
\section{System Model}
We consider a MISO system in which a transmitter equipped with $M$ antennas sends information bits to a receiver equipped with a single antenna. As illustrated in Fig. \ref{Fig1}, we assume one-bit transceiver model where the transmitter uses one-bit DACs and the receiver uses one-bit ADCs to extremely reduce the power consumption. Let the channel input vector sent by the transmitter be ${\bf \bar x}=\left[{\bar x}_1,{\bar x}_2,\ldots, {\bar x}_{M}\right]^{\top}$.  We also denote ${\bar{\bf  h}}^{\top}\in \mathbb{C}^{{1}\times {M}}$ as a frequency-flat baseband-equivalent channel between the transmitter and the receiver. The baseband-equivalent channel considered in this paper can include analog beamforming effects by considering the hybrid-beamforming architecture in \cite{Omar2014,Amed2014}. Here, the frequency-flat assumption might be valid for the millimeter-wave frequency bands in which delay-spreads are limited with analog beamforming in line of sight (LOS) channel environments \cite{Sun:12}.


When the channel input vector ${\bf \bar x}$ is sent, the complex-valued baseband received vector with one-bit ADCs is
\begin{align}
{ \bar y} = {\sf sign} \left({\bar {\bf h}}^{\top}{\bf \bar x} +{ \bar z}\right)\in \left\{-1\!-\!j,-\!1+\!j,+\!1-\!j,+\!1+\!j\right\},\label{eq:complex_inoutput} 
\end{align} 
where ${ \bar z}\in \mathbb{C}$ represents the additive noise distributed as circularly-symmetric complex Gaussian random variable with zero-mean and variance of $\sigma^2$, i.e., ${\bar z}\sim \mathcal{CN}(0,\sigma^2)$. Here, ${\sf sign}(\cdot): \mathbb{R} \rightarrow \{1,-1\}$ denotes the one-bit quantization function with ${\sf sign}(c)=1$ if $c \geq 0$ and $-1$, otherwise. This sign function is separately applied to the real and imaginary component of each received signal.

Without loss of generality, we rewrite the complex-valued input-output relationship in \eqref{eq:complex_inoutput} into an equivalent real-valued representation as
 \begin{align}
{\bf y} = {\sf sign} \left({\bf H}{\bf x}  +{\bf z}\right),\label{eq:real_inoutput}
\end{align}
where  ${\bf H}=\left[\! {\begin{array}{cc}
   {\bar {\bf h}}^{\top}_{\rm Re} &  -{\bar {\bf h}}^{\top}_{\rm Im} \\      
    {\bar {\bf h}}^{\top}_{\rm Im} &  {\bar {\bf h}}^{\top}_{\rm Re} \\
 \end{array} } \!\!\right]\in\mathbb{R}^{2 \times 2M}, \ {\bf y}=\left[{ \bar y}_{\rm Re}, {\bar y}_{\rm Im}\right]^{\!\top}\! \in\!\left\{-1,+1\right\}^{2}\!=\!\mathcal{Y}, \ {\bf x}=\left[{\bf \bar x}_{\rm Re}^{\top}, {\bf \bar x}_{\rm Im}^{\top}\!\right]^{\top}\in\mathbb{R}^{2M},$ $~{\rm and}~ \ {\bf z}=\left[{ \bar z}_{\rm Re},  { \bar z}_{\rm Im}\right]^{\top} \in \mathbb{R}^{2}$. This real-valued representation will be used in the sequel.

\subsection{Channel Input Construction}
  When information bits are encoded with one-bit DACs, each transmit antenna can only send a BPSK signal per in-phase (or quadrature) component. Unlike the previous modulation methods under the one-bit DACs constraint \cite{Gao2017,Gao2018}, we consider spatial lattice modulation (SLM) in \cite{Choi2018}, which is known as a generalized method of the spatial modulation. The key idea of the SLM is to jointly map information bits into one of $3^{2M}-1$ cubic lattice vectors in $\{-1,0,1\}^{2M}$ by using $M$ transmit antennas and the real and imaginary components per antenna. Each transmit signal of the SLM method represents a set of active dimensions and BPSK signals. Specifically, when the $m$-th component of ${\bf x}$, i.e., $x_m$ is activated, a BPSK signal $x_m\in \{-1,+1\}$ is sent. If the $m$-th component is deactivated, the zero signal is sent. Since there are ${2M \choose i}$ possible ways to select $i$ activated dimensions, a total of
\begin{align}
	\sum_{i=0}^{2M}{2M \choose i}2^i=3^{2M}
\end{align}
distinct information vectors can be generated using $M$ transmit antennas with one-bit DACs. If we discard the all-zero vector, $3^{2M}-1$ signal vectors can be used as channel inputs. We define the corresponding channel inputs of the SLM method as $\mathcal{ X}=\{-1,0,+1\}^{2M}\setminus {\bf 0}_{2M}$. In general, $\mathcal{X}$ can be regarded as a collection of spatially modulated signals with different sparsity level. With this view-point, we define some properties of $\mathcal{X}$.

\vspace{0.1cm}
{\bf Property 1 (Instantaneous transmit power):} The instantaneous transmission power (i.e., sparsity level) of any signal point in $\mathcal{X}$ belongs to $\{1,2,\ldots,2M\}$, namely, 
\begin{align}
   \|{\bf x}\|_2^2\in\{1,2,\ldots,2M\} \ \  (\forall{\bf x}\in \mathcal{X}).
\end{align}

{\bf Property 2 (Power-level signal subset):} 
We define disjoint subsets $\mathcal{X}_{u}\subset \mathcal{X}$, each of which contains vectors in $\mathcal{X}$ with the same instantaneous power level $u$, i.e., \begin{align}
	\mathcal{X}_u=\left\{{\bf x} | \|{\bf x}\|_2^2=u\right\},
\end{align}
  where $ \mathcal{X}=\bigcup_{u=1}^{2M}\mathcal{X}_{u}$. The cardinality of the $u$-th power level subset is $|\mathcal{X}_{u}|={2M \choose u}2^u$.
  
\begin{table}
\centering
	\begingroup
\caption{Look-up Table for SLM with Cubic lattices $\left( M = 2 \right)$.}
\vspace{-0.4cm}
\begin{tabular}{||c|c||c|c||}
\Xhline{3\arrayrulewidth}
 $\bf x^{\rm T}$ &  $\mathcal{X}_{u}$  & $\bf x^{\rm T}$ &  $\mathcal{X}_{u}$\\
\Xhline{3\arrayrulewidth}
\vspace{-0.05cm}${[ \pm1,   0,   0,   0 ]}$ & \multirow{4}{*}{$\mathcal{X}_{1}$}  & ${[ \pm 1,   \pm 1,   0,0]}$ & \multirow{6}{*}{$\mathcal{X}_{2}$}\\ \cline{1-1} \cline{3-3}
\vspace{-0.05cm}${[ 0,   \pm1,   0,   0 ]}$ &   & ${[ \pm 1,   0,   \pm 1,0]}$ & \\\cline{1-1} \cline{3-3}
\vspace{-0.05cm}${[ 0,   0,   \pm1,   0 ]}$ &   & ${[ \pm 1,   0,   0,\pm1]}$ & \\\cline{1-1} \cline{3-3}
\vspace{-0.05cm}${[ 0,   0,   0,   \pm1 ]}$ &   & ${[ 0,   \pm 1,  \pm 1,0]}$ & \\\cline{1-2} \cline{3-3}
\vspace{-0.05cm}${[ \pm1,   \pm1,   \pm1,   0 ]}$ & \multirow{4}{*}{$\mathcal{X}_{3}$}  & \vspace{-0.05cm}${[ 0,   \pm 1,  0,\pm 1]}$ & \\\cline{1-1} \cline{3-3}
\vspace{-0.05cm}${[ \pm1,   \pm1,   0,   \pm1 ]}$ &   & ${[ 0,   0,  \pm 1,\pm 1]}$ & \\\cline{1-1} \cline{3-4}
\vspace{-0.05cm}${[ \pm1,   0,   \pm1,   \pm1 ]}$ &   &\multirow{2}{*}{${[ \pm 1,   \pm 1,  \pm 1,\pm 1]}$} & \multirow{2}{*}{$\mathcal{X}_{4}$} \\\cline{1-1} 
\vspace{-0.05cm}${[ 0,   \pm1,   \pm1,   \pm1 ]}$ &  & & \\
\hline
\Xhline{3\arrayrulewidth}
\end{tabular} \vspace{-0.6cm}\label{table:SLM_CB}
\endgroup
\end{table}
  
{\bf Example 1 :} 
Suppose $M=2$. In this case, it is possible to generate 80 distinct channel inputs by using the SLM method. The corresponding 80 vectors are listed in Table I. Note that the channel input set generated by the conventional spatial modulation and spatial multiplexing are
\begin{align}
	{\mathcal{X}}^{\sf SM}\!=\!\left\{\!\begin{bmatrix}
    +1 \\
    0 \\ 
    +1 \\
    0 \\
\end{bmatrix},\!\begin{bmatrix}
    -1 \\
    0 \\ 
    +1 \\
    0 \\
\end{bmatrix},\!\begin{bmatrix}
    +1 \\
    0 \\ 
    -1 \\
    0 \\
\end{bmatrix},\!\begin{bmatrix}
    -1 \\
    0 \\ 
    -1 \\
    0 \\
\end{bmatrix},\! \begin{bmatrix}
    0 \\
    +1 \\ 
    0 \\
    +1 \\
\end{bmatrix},\!\begin{bmatrix}
    0 \\
    -1 \\ 
    0 \\
    +1 \\
\end{bmatrix},\!\begin{bmatrix}
   0 \\
    +1 \\ 
    0 \\
    -1 \\
\end{bmatrix},\!\begin{bmatrix}
    0 \\
    -1 \\ 
    0 \\
    -1 \\
        \end{bmatrix}\!\right\} \subset \mathcal{X}_2\nonumber
\end{align} and ${\mathcal{X}}^{\sf MG}={\mathcal{X}}_4$, respectively. Since $\mathcal{X}^{\sf SM}$ and $\mathcal{X}^{\sf MG}$ are the subsets of the SLM signal set $\mathcal{X}=\{-1,0,+1\}^{4}\setminus {\bf 0}_{4}$ with the proper real and imaginary mapping, the SLM signal set in \cite{Choi2018} generalizes the existing spatial modulation and spatial multiplexing methods in \cite{Mesleh2008,Renzo2011,Mesleh2015,Yang2015,Ibrahim2016}.

 {\bf Property 3 (Rotationally invariant signal subset in $\mathcal{X}_{u}$):} 
 We define $\mathcal{X}_{u,k}\subset \mathcal{X}_u$ as the $k$th subset of  $\mathcal{X}_u$, which contains four elements that satisfy the invariant property of the $90^{\circ}$ rotation. Since $|\mathcal{X}_{u}|={2M \choose u}2^u$, it is possible to generate $K_u={2M \choose u}2^{u-2}$ disjoint subsets $\mathcal{X}_{u,k}\subset \mathcal{X}_u$ where $k=\{1,2,\ldots, K_u\}$. Here, we say that a set $\mathcal{X}_{u,k}$ is rotationally invariant if \begin{align}
   \mathcal{X}_{u,k}= \left\{{\bf R}^0{\bf x}_{u,k},{\bf R}^1{\bf x}_{u,k},{\bf R}^2{\bf x}_{u,k},{\bf R}^3{\bf x}_{u,k}\right\}, \ \ \forall{\bf x}_{u,k}\in \mathcal{X}_{u,k},
\end{align} 
 where ${\bf R}=\begin{bmatrix}
		{\bf 0}_{M} & -	{\bf I}_{M} \\
		{\bf I}_{M} & 	{\bf 0}_{M} \\
\end{bmatrix}$ is a $90^{\circ}$ rotation matrix.  From the definition, the channel input set $\mathcal{X}$ can be decomposed with disjoint subsets $\mathcal{X}_{u,k}$ as
\begin{align}
    \mathcal{X}=\bigcup_{u=1}^{2M}\bigcup_{k=1}^{K_u}\mathcal{X}_{u,k}, \ \ \ \mathcal{X}_{u_1,k_1}\!\!\! \underset{{u_1,k_1} \neq {u_2,k_2}}{\bigcap} \!\!\!\mathcal{X}_{u_2,k_2}=\phi.
\end{align}

{\bf Example 2:} When $M=2$, we obtain two subsets of $\mathcal{X}_1$ as
\begin{align}
  {\mathcal{X}_{1,1}}= \left\{\!\begin{bmatrix}
    +1  \\
    0  \\
    0\\
    0\\
\end{bmatrix},\!\begin{bmatrix}
   -1  \\
    0  \\
    0\\
    0\\
\end{bmatrix},\!\begin{bmatrix}
   0  \\
    0  \\
    +1\\
    0\\
\end{bmatrix},\!\begin{bmatrix}
    0  \\
    0  \\
    -1\\
    0\\
\end{bmatrix}\!\right\}, \ \ 
{\mathcal{X}_{1,2}}=\left\{\!\begin{bmatrix}
    0  \\
    +1  \\
    0\\
    0\\
\end{bmatrix},\!\begin{bmatrix}
   0  \\
    -1  \\
    0\\
    0\\
\end{bmatrix},\!\begin{bmatrix}
   0  \\
    0  \\
    0\\
    +1\\
\end{bmatrix},\!\begin{bmatrix}
    0  \\
    0  \\
    0\\
    -1\\
\end{bmatrix}\!\right\}. 
\end{align}
In a complex-valued representation, $\mathcal{X}_{1,1}$ and $\mathcal{X}_{1,2}$ can be equivalently written as
\begin{align}
  {\bar{\mathcal{X}}_{1,1}}=\left\{\!\begin{bmatrix}
    +1  \\
    0  \\
\end{bmatrix},\!\begin{bmatrix}
   -1  \\
    0  \\
\end{bmatrix},\!\begin{bmatrix}
  +j  \\
    0  \\
\end{bmatrix},\!\begin{bmatrix}
 -j  \\
    0  \\
\end{bmatrix}\!\right\}, \ \ 
	{\bar{\mathcal{X}}_{1,2}}=\left\{\!\begin{bmatrix}
0 \\
    +1  \\
\end{bmatrix},\!\begin{bmatrix}
 0 \\
    -1  \\
\end{bmatrix},\!\begin{bmatrix}
 0 \\
    +j  \\
\end{bmatrix},\!\begin{bmatrix}
0 \\
    -j  \\
\end{bmatrix}\!\right\}. 
\end{align}
In a spatial domain, the rotationally invariant property of ${\mathcal{X}}_{u,k}$ implies that the elements in ${\mathcal{X}}_{u,k}$ share the same antenna activation pattern.

 {\bf Property 4 (The average transmission power):}  For each $\mathcal{X}_{u,k}$, we define the corresponding probability mass function as $p_{u,k}={\rm Pr}\left[{\bf x} \in \mathcal{X}_{u,k}\right]$. Then, the average transmit power is
 \begin{align} 
\mathbb{E}\left[\|{\bf x}\|_{2}^{2}\right]=\sum_{u=1}^{2M}u\sum_{k=1}^{K_u}p_{u,k}\leq P_{\rm t},
\end{align} 
where $P_{\rm t}$ is the average power constraint in the system.  Throughout this paper, we define the signal-to-noise ratio (SNR) as
\begin{align}
    \text{SNR}=\frac{P_{\rm t}}{\sigma^2}.
\end{align}

\vspace{-0.1cm}
\subsection{Channel Capacity}
In this paper, we consider the channel capacity when the perfect knowledge of CSIT and CSIR are available at the transceiver. When the channel is fixed over the duration of spanning codewords, the capacity of the constant MISO channel with one-bit ADCs and DACs is obtained by solving the following optimization problem:
\begin{align}
   & \ \ \ \ C= \max_{{\rm Pr}[{\bf x}]}  {{\sf I}\left({\bf x};{\bf y}|{\bf H}=\mathcal{H} \right)}, \  \mathcal{H}=\begin{bmatrix}
   {\bf h}_1^{\top}  \\
    {\bf h}_2^{\top}  \\
\end{bmatrix}, \label{eq:Capacity_1}\nonumber\\
 \textrm{s.t.}& \ \mathbb{E}\left[\|{\bf x}\|_2^2\right]\leq P_{\rm t},  \ {\bf x}\in \mathcal{ X}=\{-1,0,+1\}^{2M}\setminus {\bf 0}_{2M},
\end{align}
where 
\begin{align}
	&{\sf I}\!\left({\bf x};{\bf y}\right)	=\!\sum_{{\bf x}\in{\mathcal{X}}}\!\sum_{{\bf y}\in\mathcal{Y}}\!{\rm Pr}[{\bf x}]{\rm Pr}[{\bf y}| {\bf x}]\log_2\!\!\frac{{\rm Pr}[{\bf y}| {\bf x}]}{{\rm Pr}[{\bf y}]}\!.\label{eq:MI_fixedchannel}
\end{align}
Our goal in this paper is to find the closed form expression of the capacity by deriving the optimal input distribution over $\mathcal{X}$.

\vspace{-0.1cm}
\section{SISO Channel with One-Bit Transceiver}
 In this section, we characterize the fundamental limit of the SISO channel when one-bit ADCs and DACs are employed at the transceiver. Throughout this section, we omit the index $k$ for the notational simplicity, i.e., $\mathcal{X}_{u}=\mathcal{X}_{u,1}$ and ${\bf x}_{u,1}={\bf x}_u$, because there is an unique subset $\mathcal{X}_{u,k}$ in each power level set $\mathcal{X}_u$.
 
\vspace{-0.1cm}
 \subsection{Capacity of the SISO Channel}
 
The following theorem is the main result of this section. 
 
 \vspace{0.2cm}
 {\bf Theorem 1:}   	For a given channel realization, i.e., ${\bf H}={\mathcal{H}}$, let ${\sf H}_b^{\mathcal{X}_{u}}
$ be the sum of binary entropy functions for different power level input subsets, i.e.,
\begin{align}
	 {\sf H}_b^{\mathcal{X}_{u}}=\sum_{n=1}^{2}{\sf H}_b\left(Q\left(\sqrt{\frac{2}{\sigma^2}}{\bf h}_n^{\top}{\bf x}_{u}\right)\right),
\end{align} 
	where ${\sf H}_b(x)=-x\log_2x-(1-x)\log_2(1-x)$ for $0<x<1$. Then, the capacity of the SISO channel with one-bit ADCs and DACs is
{{\begin{align}
	&C^{\sf SISO}\!=\!\begin{cases}
    2-\!{\sf{H}}_b^{\mathcal{X}_{1}},   \!\!&\!\!\!\text{if } \left\{{\sf H}_b^{\mathcal{X}_{1}}\! \leq \!{\sf H}_b^{\mathcal{X}_{2}}\right\}\!\cup\!\{P_{\rm t}\!=\!1\},\\
  2\!-\!(2\!-\!P_{\rm t}){\sf{H}}_b^{\mathcal{X}_{1}}\!-(P_{\rm t}\!-\!1){\sf{H}}_b^{\mathcal{X}_2}, \!\!&\!\!\! \text{if } \left\{{\sf H}_b^{\mathcal{X}_{1}}\!>\! {\sf H}_b^{\mathcal{X}_{2}} \right\}\!\cap\! \{1\!<\!P_{\rm t}\!<\! 2\},\\
          2\!-\!{\sf{H}}_b^{\mathcal{X}_{2}}, \!\!&\!\!\!\text{if } \left\{{\sf H}_b^{\mathcal{X}_1}\!> \!{\sf H}_b^{\mathcal{X}_{2}}\right\}\!\cap\!\{P_{\rm t}\!=\!  2\}.     \end{cases} \label{eq:Theorem1}
\end{align}}}
\vspace{-0.2cm}
\begin{IEEEproof}
The proof consists of two steps. We first specify the property of the optimal input distribution, which is stated in the following lemma:

\vspace{0.1cm}
{\bf Lemma 1:} For the SISO channel with one-bit ADCs/DACs, there exists a capacity-achieving input distribution which is uniform in each $\mathcal{X}_u$, namely,
\begin{align}
	{\rm Pr}\left[{\bf x}={\bf R}^i{\bf x}_u\right]=\frac{p_u}{4}, \ \forall i \in \{0,1,2,3\},
\end{align} for $u \in \{1,2\}.$
\begin{IEEEproof}
See Appendix A.
\end{IEEEproof}
By leveraging Lemma 1, we derive the capacity-achieving input distribution by finding the optimal probabilities $p_1$ and $p_2$ so as to maximize the mutual information under the average power constraint. 

We start by rewriting the input-output relationship in \eqref{eq:real_inoutput} for the SISO channel as
\begin{align}
{\bf y}=\begin{bmatrix}
	{y}_{1} \\
	{y}_{2}
\end{bmatrix}=
	\begin{bmatrix}
	{\sf sign}\left({\bf h}_1^{\top}{\bf x}+{z}_{1}\right) \\
	{\sf sign}\left({\bf h}_2^{\top}{\bf x}+{z}_{2}\right)
\end{bmatrix}.
\end{align}
The mutual information between the input ${\bf x}$ and the output ${\bf y}$ for a given channel $\mathcal{H}$ is 
\begin{align}
	&{\sf I}({\bf x};{\bf y}|{\bf H}=\mathcal{H})={\sf H}({\bf y}|{\bf H}=\mathcal{H}) - {\sf H}({\bf y}|{\bf x},{\bf H}=\mathcal{H}) \label{eq:MI_SISO}.
\end{align}
To compute ${\sf H}({\bf y}|{\bf H}=\mathcal{H})$ in \eqref{eq:MI_SISO}, we calculate the conditional probability of ${\bf y}$ when the channel is given as ${\bf H}=\mathcal{H}$. We first compute the conditional probability of ${\bf y}=[1,1]^{\top}$, which is
\begin{align}
{\rm Pr}\left[{\bf y}=[1,1]^{\top}| \ {\bf H}=\mathcal{H}\right]&=\sum_{u=1}^{2}\sum_{i=0}^{3} {\rm Pr}\left[{\bf y}=[1,1]^{\top}| \ {\bf x}\!=\!{\bf R}^i{\bf x}_{u}, {\bf H}=\mathcal{H}\right]{\rm Pr}\left[{\bf x}\!=\!{\bf R}^i{\bf x}_{u}\right]  \nonumber \\
&\stackrel{(a)}{=}\sum_{u=1}^{2}\sum_{i=0}^{3}\left(\prod_{n=1}^{2} {\rm Pr}\left[{ y}_n=1|{\bf x}\!=\!{\bf R}^i{\bf x}_{u},{\bf h}_n^{\top} \right]\right){\rm Pr}\left[{\bf x}\!=\!{\bf R}^i{\bf x}_{u}\right]\nonumber\\
&\stackrel{(b)}{=}\sum_{u=1}^{2}\frac{p_u}{4}\sum_{i=0}^{3}\prod_{n=1}^{2} {\rm Pr}\left[y_n=1|{\bf x}\!=\!{\bf R}^i{\bf x}_{u},{\bf h}_n^{\top}\right], \label{eq:cond_entropy}
\end{align} 
where (a) follows from the conditional independence of $y_1$ and $y_2$ for the given channel and input vectors, and (b) is due to the result from Lemma 1. By using the following identities, i.e.,
\begin{align}
    {\bf h}_{1}^{\top}{\bf R}=-{\bf h}_{2}^{\top} ~~{\rm and}~~ {\bf h}_{2}^{\top}{\bf R}={\bf h}_{1}^{\top}, \label{eq:relation}
\end{align}
we further simplify
\begin{align}
\sum_{i=0}^{3}\prod_{n=1}^{2} {\rm Pr}\left[y_n=1|{\bf x}\!=\!{\bf R}^i{\bf x}_{u},{\bf h}_n^{\top}\right]
	&\!=\!{\rm Pr}[z_1\!>\!-{\bf h}_1^{\top}{\bf x}_u]{\rm Pr}[z_2\!>\!-{\bf h}_2^{\top}{\bf x}_u]\!+\!{\rm Pr}[z_1\!>\!{\bf h}_2^{\top}{\bf x}_u]{\rm Pr}[z_2\!>\!-{\bf h}_1^{\top}{\bf x}_u]\nonumber\\&+\!{\rm Pr}[z_1\!>\!{\bf h}_1^{\top}{\bf x}_u]{\rm Pr}[z_2\!>\!{\bf h}_2^{\top}{\bf x}_u]\!+\!{\rm Pr}[z_1\!>\!-{\bf h}_2^{\top}{\bf x}_u]{\rm Pr}[z_2\!>\!{\bf h}_1^{\top}{\bf x}_u]
	\stackrel{(a)}=1. \label{eq:expand}
\end{align}
where (a) follows from ${\rm Pr}[z>x]+{\rm Pr}[z>-x]=1$. By plugging \eqref{eq:expand} into \eqref{eq:cond_entropy}, we obtain
\begin{align}
	{\rm Pr}\left[{\bf y}=[1,1]^{\top}| \ {\bf H}=\mathcal{H}\right]=\sum_{u=1}^{2}\frac{p_u}{4}=\frac{1}{4}. \label{eq:equiprobable}
\end{align}
From \eqref{eq:equiprobable} and the symmetry of the channel inputs, we conclude that the channel outputs are uniformly distributed regardless of the channel realizations. As a result, the channel output entropy is
	\begin{align}
		{\sf H}({\bf y}|{\bf H}=\mathcal{H})=2.
	\end{align}

We also compute the conditional entropy ${\sf H}\left({\bf y}|{\bf x},{\bf H}=\mathcal{H}\right)$ in \eqref{eq:MI_SISO}. From the independence of $z_1$ and $z_2$, we obtain
\begin{align}
    {\sf H}\left({\bf y}|{\bf x},{\bf H}=\mathcal{H}\right)= {\sf H}(y_1|{\bf x},{\bf h}_1^{\top}) +{\sf H}(y_2|{\bf x},{\bf h}_2^{\top}).
\end{align}
Then, the conditional entropy is computed as
\begin{align}
	{\sf H}\left({\bf y}|{\bf x},{\bf H}=\mathcal{H}\right)
	&\stackrel{(a)}=\!\sum_{u=1}^{2}\frac{p_u}{4}\!\sum_{i=0}^{3}\left\{{\sf H}\!\left(y_1|{\bf x}\!=\!{\bf R}^i{\bf x}_{u},{\bf h}_{1}^{\top}\right)+{\sf H}\!\left(y_2|{\bf x}\!=\!{\bf R}^i{\bf x}_{u},{\bf h}_{2}^{\top}\right)\right\} \nonumber\\
	&\stackrel{(b)}{=}\!\sum_{u=1}^{2}p_u\left\{{\sf H}\!\left(y_1|{\bf x}\!=\!{\bf x}_{u},{\bf h}_{1}^{\top}\right)+{\sf H}\!\left(y_2|{\bf x}\!=\!{\bf x}_{u},{\bf h}_{2}^{\top}\right)\right\},
\end{align}
where (a) follows from Lemma 1 and (b) comes from the identities in \eqref{eq:relation}.
To calculate ${\sf H}\!\left(y_n|{\bf x}\!=\!{\bf x}_{u},{\bf h}_{n}^{\top}\right)$, we compute the conditional probability of $y_n$ as
	\begin{align}
	{\rm Pr}[ y_n=1|{\bf x}={\bf x}_u, {\bf h}_n^{\top}]={\rm Pr}[ z_n > -{\bf h}_n^{\top}{\bf x}_{u}]=\!1\!-\!Q\left(\!\sqrt{\frac{2}{\sigma^2}}{\bf h}_n^{\top}{\bf x}_{u}\!\right), \label{eq:subchannel}
\end{align}
where $Q(x)=\int_{x}^{\infty}\frac{1}{\sqrt{2\pi}}\exp({-\frac{t^2}{2}})dt$ is the tail probability of the standard Gaussian distribution. Using \eqref{eq:subchannel}, the conditional entropy in \eqref{eq:MI_SISO} boils down to
\begin{align}
	&{\sf H}\left({\bf y}|{\bf x},{\bf H}=\mathcal{H}\right)\stackrel{(a)}{=}\sum_{u=1}^{2}p_u\!\sum_{n=1}^{2}{\sf H}_b\left(Q\left(\sqrt{\frac{2}{\sigma^2}}{\bf h}_n^{\top}{\bf x}_{u}\right)\right),
\end{align}
where (a) is due to the symmetry property of the binary entropy function, i.e., ${\sf H}_b(x)={\sf H}_b(1-x)$. As a result, the mutual information of the SISO channel with one-bit ADCs and DACs is
\begin{align}
	{\sf I}({\bf x};{\bf y}|{\bf H}=\mathcal{H})&={\sf H}({\bf y}|{\bf H}=\mathcal{H}) - \sum_{n=1}^2 {\sf H}\left(y_n|{\bf x},{\bf H}=\mathcal{H}\right)  \nonumber\\
	&=2-\sum_{u=1}^{2}p_u\!\sum_{n=1}^{2}{\sf H}_b\left(Q\left(\sqrt{\frac{2}{\sigma^2}}{\bf h}_n^{\top}{\bf x}_{u}\right)\right)=2-{p_1} {\sf{H}}_b^{\mathcal{X}_1}-{p_2}{\sf{H}}_b^{\mathcal{X}_2}.\label{eq:MI_SISO2} 
\end{align}

To find the capacity, we need to optimize the input distribution, i.e., $p_1$ and $p_2$. From \eqref{eq:MI_SISO2}, the maximization of ${\sf I}({\bf x};{\bf y}|{\bf H}=\mathcal{H})$ with respect to $p_1$ and $p_2$ under the average power constraint $p_1+2p_2\leq P_{\rm t}$, is equivalent to solve the following linear programming problem:
\begin{align}
   &\min_{p_1,p_2}~~  p_1{\sf H}_b^{\mathcal{X}_1}+p_2{\sf H}_b^{\mathcal{X}_2}\label{eq:Capacity_1_LP}\nonumber\\
 &~\textrm{such that}~~p_1+2p_2\leq P_{\rm t}, \nonumber\\
  &~~~~~~~~~~~~~~p_1+p_2=1, \nonumber\\
 &~~~~~~~~~~~~~~p_1\geq 0~ {\rm and} ~ p_2\geq 0.
\end{align}
Using the standard simplex method, we obtain a closed form solution of the capacity-achieving input distribution as
\begin{align}
	(p_1^{\star},p_2^{\star})\!=\!\!
\begin{cases}
    \left(1,0\right),    &\!\!\! \text{if } \left\{{\sf H}_b^{\mathcal{X}_1}\! \leq \!{\sf H}_b^{\mathcal{X}_2}\right\}\cup\{P_{\rm t}\!=\!1\},\\
   \left({2-P_{\rm t}},{P_{\rm t}-1}\right),    &\!\!\! \text{if } \left\{{\sf H}_b^{\mathcal{X}_1}\!>\! {\sf H}_b^{\mathcal{X}_2} \right\}\cap \{1\!<\!P_{\rm t}\!<\! 2\},\\
   \left(0,1\right), &\!\!\!  \text{if } \left\{{\sf H}_b^{\mathcal{X}_1}\!> \!{\sf H}_b^{\mathcal{X}_2}\right\}\cap\{P_{\rm t}\!=\!2\}.
     \end{cases} \label{eq:optimal_input}
\end{align}
By invoking \eqref{eq:optimal_input} into \eqref{eq:MI_SISO2}, the capacity of the SISO channel with one-bit ADCs and DACs is given as in \eqref{eq:Theorem1}. This completes the proof.
\end{IEEEproof}

\vspace{-0.2cm}
\subsection{Implication}

To shed further light on the importance of Theorem 1, we explain the capacity expression for three cases.   

{\bf Case 1:} When $\left\{{\sf H}_b^{\mathcal{X}_1}\! \leq \!{\sf H}_b^{\mathcal{X}_2}\right\}$, the capacity-achieving transmission strategy is to use the spatial modulation method, i.e., $\mathcal{X}_1$ uniformly. This is because ${\sf H}_b^{\mathcal{X}_1}\! \leq \!{\sf H}_b^{\mathcal{X}_2}$ implies that the effect of the phase alignment between the inputs and the channel is more important than the effect of transmission power. 



In addition, in the case of $P_{\rm t}=1$, the optimal transmission method is also to uniformly use the input signals in $\mathcal{X}_1$ regardless of channel realizations. The reason is that the use of input signals in $\mathcal{X}_2$ is infeasible to satisfy the average power constraint. In these cases, the channel capacity expression in Theorem 1 boils down to
\begin{align}
	C^{\sf SISO}&=2-\sum_{n=1}^{2}{\sf H}_b\left(Q\left(\sqrt{\frac{2}{\sigma^2}}{\bf h}_n^{\top}{\bf x}_{1}\right)\right)=2-{\sf H}_b\left(Q\left(\sqrt{\frac{2|{\bar h}_{\rm Re}|^2}{\sigma^2}}\right)\right)-{\sf H}_b\left(Q\left(\sqrt{\frac{2|{\bar h}_{\rm Im}|^2}{\sigma^2}}\right)\right). 
	\end{align}
	
\begin{figure}
\centerline{\includegraphics[width=6cm]{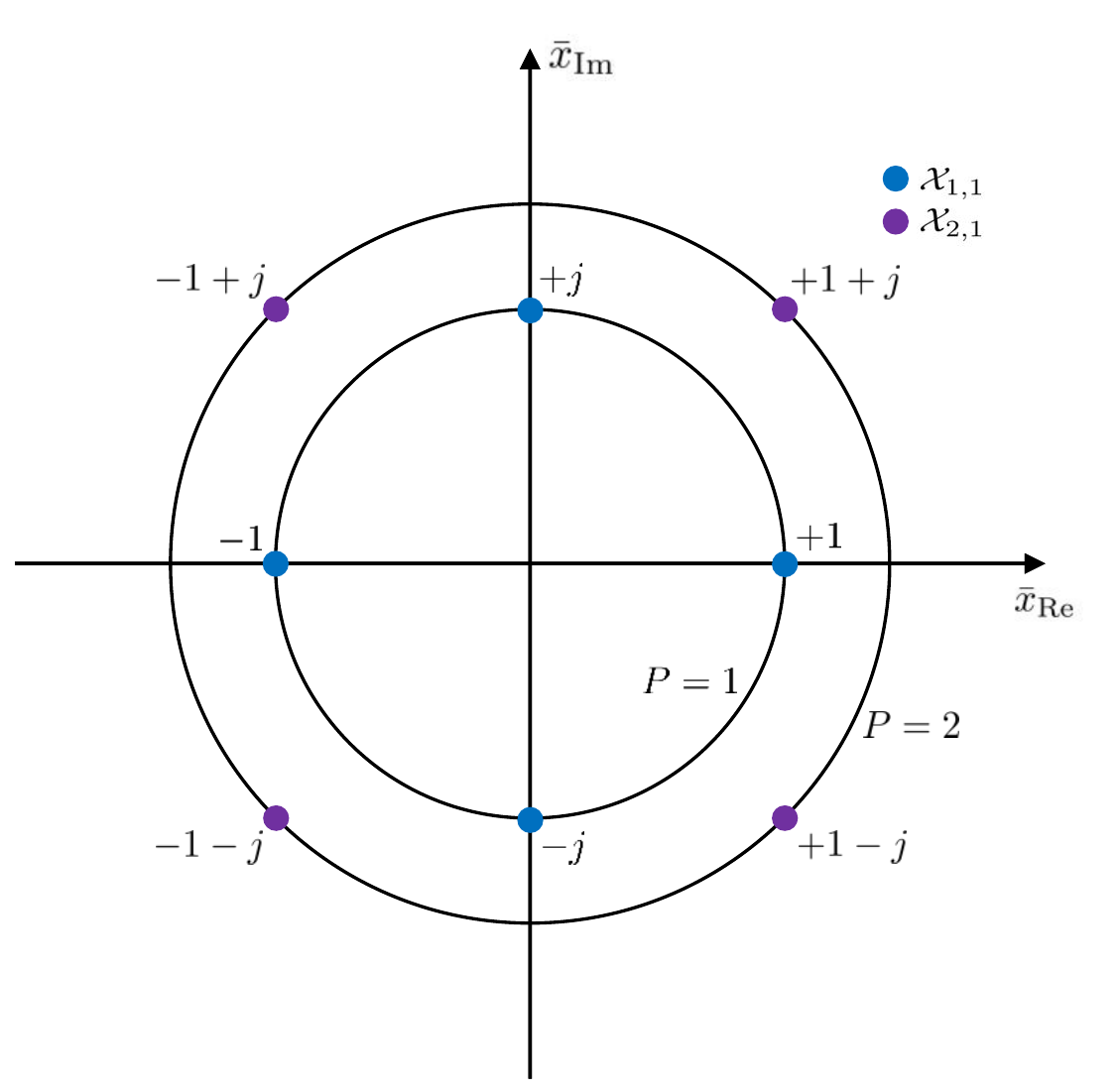}}\vspace{-0.3cm}
\caption{The eight possible channel input vectors for a single-antenna transmitter using one-bit DACs.}
\label{Fig2}
\vspace{-0.5cm}
\end{figure}

{\bf Case 2:} When $\left\{{\sf H}_b^{\mathcal{X}_1}\!>\! {\sf H}_b^{\mathcal{X}_2} \right\}$ and  $\{1\!<\!P_{\rm t}\!<\! 2\}$, the capacity-achieving transmission method is to use all signal points generated by the spatial lattice modulation method. Here, the signal points in $\mathcal{X}_1$ and $\mathcal{X}_2$ are used with the probability of $2-P_{\rm t}$ and $P_{\rm t}-1$, respectively. To provide an intuition on the design of the practical communication scheme, we can equivalently rewrite the mutual information expression in \eqref{eq:MI_SISO2} as
\begin{align}
    	{\sf I}({\bf x};{\bf y}|{\bf H}=\mathcal{H})&=2-{p_1} {\sf{H}}_b^{\mathcal{X}_1}-{p_2}{\sf{H}}_b^{\mathcal{X}_2}=p_1(2-{\sf{H}}_b^{\mathcal{X}_1})+p_2(2-{\sf{H}}_b^{\mathcal{X}_2}).
    	\label{eq:MI_SISO3} 
\end{align}
From \eqref{eq:MI_SISO3}, one can achieve the same transmission rates by time sharing between the spatial modulation method and the QPSK signaling with the time fractions of $p_1$ and $p_2$.

Since the channel is better matched with the input signals in $\mathcal{X}_2$ than in $\mathcal{X}_1$, the transmitter first harnesses $\mathcal{X}_2$ uniformly during the ${P_{\rm t}-1}$ fractions of the entire channel uses. Since $P_{\rm t}\!<\! 2$, it is impossible to use the inputs in $\mathcal{X}_2$ for all the channel uses to satisfy the average power constraint. Therefore, for the remaining fractions of the time, i.e., $2-P_{\rm t}$, it exploits the inputs in $\mathcal{X}_1$. For this case, the capacity expression is
 \begin{align}
	C^{\sf SISO}&=2 - 
	(2\!-\!P_{\rm t})\!\sum_{n=1}^2\!{\sf H}_b\small{\left(\!Q\!\left(\!\sqrt{\frac{2}{\sigma^2}}{\bf h}_n^{\top}{\bf x}_1\!\right)\right)} -(P_{\rm t}\!-\!1)\!\sum_{n=1}^{2}\!{\sf H}_b\left(\!Q\!\left(\!\sqrt{\frac{2}{\sigma^2}}{\bf h}_n^{\top}{\bf x}_2\!\right)\right)\nonumber\\
	&=2-\left(2-P_{\rm t}\right)\left[{\sf H}_b\left(Q\left(\sqrt{\frac{2|{\bar h}_{\rm Re}|^2}{\sigma^2}}\right)\right)+{\sf H}_b\left(Q\left(\sqrt{\frac{2|{\bar h}_{\rm Im}|^2}{\sigma^2}}\right)\right)\right]\nonumber\\
	&~{-\left(P_{\rm t}\!-\!1\right)\left[\!{\sf H}_b\left(\!Q\!\left(\!\sqrt{\frac{2({\bar h}_{\rm Re}\!+\!{\bar h}_{\rm Im})^2}{\sigma^2}}\!\right)\!\right)\!+\!{\sf H}_b\left(\!Q\!\left(\!\sqrt{\frac{2({\bar h}_{\rm Re}\!-\!{\bar h}_{\rm Im})^2}{\sigma^2} }\!\right)\!\right)\right]\!}.
	\end{align}

{\bf Case 3:} For the case of $\left\{{\sf H}_b^{\mathcal{X}_1}\!>\! {\sf H}_b^{\mathcal{X}_2} \right\}$ and $\{ P_{\rm t}\!=\! 2\}$, the uniform QPSK signaling achieves the capacity because the input vectors in $\mathcal{X}_2$ are better aligned with the channel vector than those in $\mathcal{X}_1$, and the transmission power is sufficient to meet the average power constraint. In this case, the capacity becomes	
\begin{align}
	C^{\sf SISO}&= 2-\sum_{n=1}^2{\sf H}_b\left(Q\left(\sqrt{\frac{2}{\sigma^2}}{\bf h}_n^{\top}{\bf x}_2\right)\right)=2\!-\!{\sf H}_b\left(Q\!\left(\!\sqrt{\frac{2({\bar h}_{\rm Re}\!+\!{\bar h}_{\rm Im})^2}{\sigma^2}}\right)\right)\!-\!{\sf H}_b\left(Q\!\left(\!\sqrt{\frac{2({\bar h}_{\rm Re}\!-\!{\bar h}_{\rm Im})^2}{\sigma^2}} \right)\right). 
	\end{align} 
\begin{figure}
\centerline{\includegraphics[width=8cm]{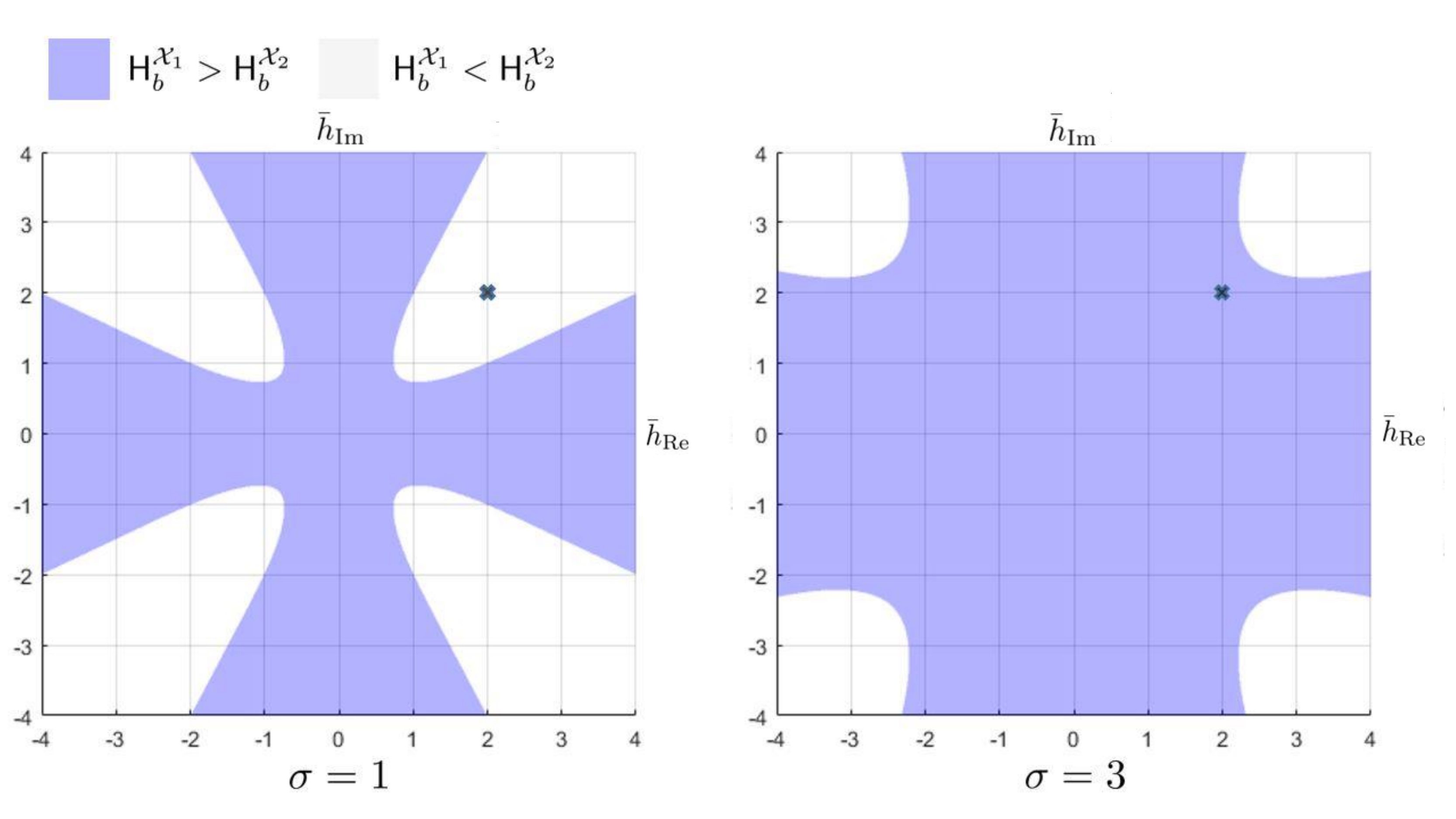}}\vspace{-0.3cm}
\caption{The channel capacity comparison when $M=1$ as a function of the channel phase, magnitude, and the SNR when using two different channel input sets, i.e., $\mathcal{X}_1$ and $\mathcal{X}_2$.}
\label{Fig3}\vspace{-0.5cm}
\end{figure}
	{\bf Remark 2 (Transmission strategy according to the SNR):}
	For the SISO fading channel, the region of the channel realizations satisfying the condition of ${\sf{H}}_b^{\mathcal{X}_1}\!>\! {\sf{H}}_b^{\mathcal{X}_2}$ is depicted in Fig. \ref{Fig3}. This figure elucidates that the optimal transmission strategy depends on the SNR. For example, when we set $h=2+2j$ and the noise variance $\sigma^2=1$, the spatial modulation method, i.e., $\mathcal{X}_1$ achieves a higher spectral efficiency than the QPSK signaling, i.e., $\mathcal{X}_2$. Whereas, at the lower SNR, i.e., $\sigma^2=9$, the use of QPSK signaling achieves a higher spectral efficiency than the spatial modulation method. As the SNR increases, the region of the channel values, where the use of the spatial modulation method is the optimal, is also expanded. This implies that the phase alignment between the channel and the inputs becomes more dominant in the capacity than the instantaneous transmission power at the high SNR regime.

		{\bf Corollary 1:} One-bit CSIT achieves the capacity of the SISO channel with one-bit ADCs and DACs.
	\begin{IEEEproof}
	The proof is evident from the interpretation of Theorem 1. To achieve the capacity, the receiver needs to send back one-bit feedback information that indicates whether to use the spatial modulation method or not. Since the transmitter knows the average power constraint $P_{\rm t}$, it is possible to use the capacity-achieving transmission strategy with the one-bit feedback information.
	\end{IEEEproof}
	
		{\bf Corollary 2:} Without CSIT, the capacity using one-bit ADCs and DACs is
		\begin{align}
			C_{\sf CSIR}^{\sf SISO}\!\!=\!\!\begin{cases}
			2\!-\!{\sf{H}}_b^{\mathcal{X}_1}, \!\!&\!\!\text{if } \{P_{\rm t}\! =1\}\\
    2\!-\!(2-P_{\rm t}){\sf{H}}_b^{\mathcal{X}_1}\!-\!(P_{\rm t}-1){\sf{H}}_b^{\mathcal{X}_2},   \!\!&\!\!\text{if } \{1<P_{\rm t}<2\},\\
          2\!-\!{\sf{H}}_b^{\mathcal{X}_2}, \!\!&\!\!\text{if } \{P_{\rm t}\! = 2\}.    \end{cases}
		\end{align}
		\begin{IEEEproof}
Since the transmitter has no information whether the channel is better aligned to ${\mathcal{X}_1}$ or ${\mathcal{X}_2}$, the optimal strategy is to transmit $\mathcal{X}_2$ as much as possible, which uses more instananous power than $\mathcal{X}_1$.
\end{IEEEproof}
The capacity loss due to the lack of CSIT is, therefore, 
\begin{align}
	\Delta C_{\sf CSIT}^{\sf SISO} \!=\!\begin{cases}
    (P_{\rm t}-1)\left({\sf{H}}_b^{\mathcal{X}_2}\!-\!{\sf{H}}_b^{\mathcal{X}_1}\right),   \!\!&\!\!\text{if }\! \left\{{\sf H}_b^{\mathcal{X}_1}\!<\! {\sf H}_b^{\mathcal{X}_2} \right\},\\
          0 \ ,\!\!&\!\!\text{if }\!\left\{{\sf H}_b^{\mathcal{X}_1}\!>\! {\sf H}_b^{\mathcal{X}_2} \right\}.    \end{cases} \label{eq:gap}
\end{align}
In Fig. \ref{Fig3}, it has been shown that the portion of ${\sf H}_b^{\mathcal{X}_1}\!<\! {\sf H}_b^{\mathcal{X}_2}$ increases with the SNR. From \eqref{eq:gap}, the capacity gap due to the lack of CSIT, i.e., $\Delta C_{\sf CSIT}^{\sf SISO}$ also increases. In the low SNR regime, however, the loss disappears; this implies that the impact of CSIT is negligible.

\vspace{-0.1cm}
\subsection{Capacity Loss Analysis by One-Bit DACs} 

In this subsection, we characterize the capacity loss by the use of one-bit DACs at the transmitter. To accomplish this, we compare our capacity result in Theorem 1 with the result in \cite{Mo2015_TSP}, where infinite-precision DACs are used at the transmitter.

When infinite-precision DACs are employed, the SISO capacity expression with the perfect CSIT and CSIR is derived in \cite{Mo2015_TSP}. The capacity-achieving input is the rotated QPSK, namely, 
\begin{align}
{\mathcal{X}}_{\sf{ADCs}}=\bigg\{ \small{\frac{1}{\sqrt{{\bar h}_{\rm Re}^2+{\bar h}_{\rm Im}^2}}}\left[ {\begin{array}{cc}
   {\bar h}_{\rm Re} &  {\bar h}_{\rm Im} \\      
    -{\bar h}_{\rm Im} &  {\bar h}_{\rm Re} \\
 \end{array} } \right]    {\bf x} {\mid} {\bf x}\in \mathcal{X}_2\bigg\}.
\end{align}
By perfectly align the channel inputs to the channel direction, the channel can be decoupled into two real-valued sub-channels, each with the same channel gain. For the decoupled channels, the BPSK signaling is shown to be optimal, which leads to the the simple capacity expression as
\begin{align}
	C_{\sf ADCs}^{\sf SISO}=2\left(1-{\sf H}_b\left(Q\!\left(\sqrt{\frac{2}{\sigma^2}({\bar h}_{\rm Re}^2+{\bar h}_{\rm Im}^2)}\!\right)\right)\right). \label{eq:CSIT_SISO}
\end{align}
By using \eqref{eq:Theorem1} and \eqref{eq:CSIT_SISO}, the capacity loss by the use of one-bit DACs for the SISO channel when $P_{\rm t}=2$ is characterized as in the following corollary. 

{\bf Corollary 3:}  The capacity loss $\Delta C_{\sf DACs}^{\sf SISO}=C_{\sf ADCs}^{\sf SISO}-C^{\sf SISO}$ is 
\begin{align}
	\Delta C_{\sf DACs}^{\sf SISO}&=\!\begin{cases}
   {\sf{H}}_b^{\mathcal{X}_1}-2{\sf H}_b\!\left(Q\!\left(\!\sqrt{\frac{2}{\sigma^2}({\bar h}_{\rm Re}^2\!+\!{\bar h}_{\rm Im}^2)}\right)\!\right),   \!\!&\!\!\text{if } \left\{{\sf H}_b^{\mathcal{X}_1}\!<\! {\sf H}_b^{\mathcal{X}_2} \right\},\\
          {\sf{H}}_b^{\mathcal{X}_2}-2{\sf H}_b\!\left(Q\!\left(\!\sqrt{\frac{2}{\sigma^2}({\bar h}_{\rm Re}^2\!+\!{\bar h}_{\rm Im}^2)}\right)\!\right), \!\!&\!\!\text{if }\left\{{\sf H}_b^{\mathcal{X}_1}\!>\! {\sf H}_b^{\mathcal{X}_2} \right\}.    \end{cases} \label{eq:gapdac}
 \end{align}
 \begin{figure}
\centerline{\includegraphics[width=6cm]{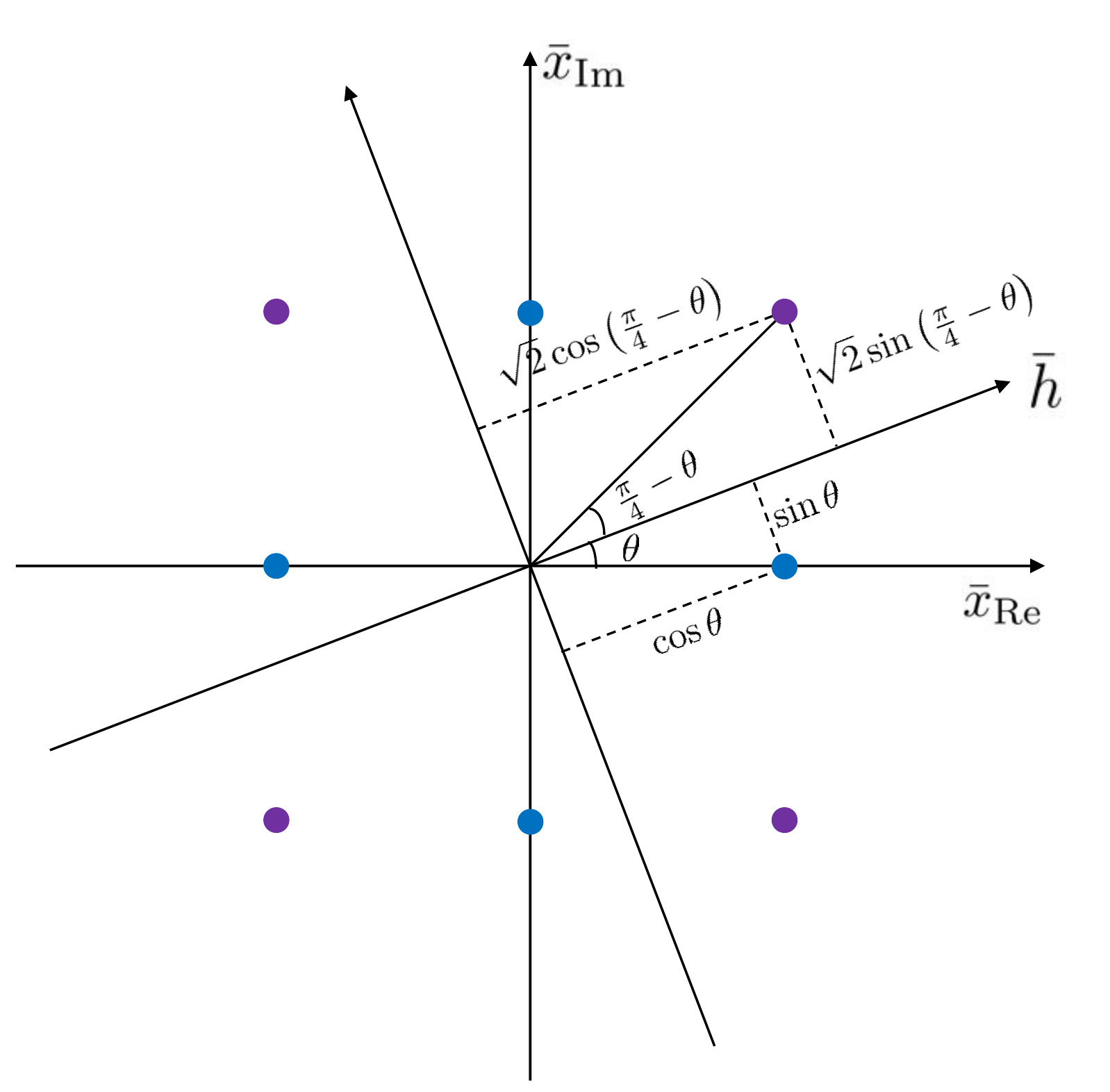}}\vspace{-0.3cm}
\caption{Phase misalignment between the channel and the constellation points.}
\label{Fig4}\vspace{-0.5cm}
\end{figure}
To provide an intuition, we derive an upper bound of the loss in \eqref{eq:gapdac}. As illustrated in Fig. \ref{Fig4}, the capacity of the SISO channel with one-bit ADCs and DACs can be equivalently rewritten as
 \begin{align}
 C&^{\sf{SISO}}=\max\Bigg\{ 2-{\sf H}_b\left(Q\left(\sqrt{\frac{2}{\sigma^2}({\bar h}_{\rm Re}^2\!+\!{\bar h}_{\rm Im}^2)\cos^2\theta}\right)\right)-{\sf H}_b\left(Q\left(\sqrt{\frac{2}{\sigma^2}({\bar h}_{\rm Re}^2\!+\!{\bar h}_{\rm Im}^2)\sin^2\theta}\right)\right),\nonumber\\&2-{\sf H}_b\left(Q\left(\sqrt{\frac{4}{\sigma^2}({\bar h}_{\rm Re}^2\!+\!{\bar h}_{\rm Im}^2)\cos^2\left(\frac{\pi}{4}-\theta\right)}\right)\right)-{\sf H}_b\left(Q\left(\sqrt{\frac{4}{\sigma^2}({\bar h}_{\rm Re}^2\!+\!{\bar h}_{\rm Im}^2)\sin^2\left(\frac{\pi}{4}-\theta\right)}\right)\right)
 \Bigg\}, \label{eq:SISO_rerepresent}
 \end{align}
 where $\theta$ denotes the phase of $\bar{h}$ \cite{Mo2016}. To derive an upper bound of \eqref{eq:gapdac}, we derive a lower bound of $C^{\sf SISO}$ by considering a suboptimal strategy which exploits $\mathcal{X}_1$ or $\mathcal{X}_2$ by simply observing ${\theta}$, i.e.,
 \begin{align}
     \begin{cases} \mathcal{X}_1, \ \  \text{if } \ \sqrt{2}\sin\left(\frac{\pi}{4}-\theta\right)<\sin\theta,\\
     \mathcal{X}_2, \ \  \text{if } \ \sqrt{2}\sin\left(\frac{\pi}{4}-\theta\right)>\sin\theta. \label{eq:threshold}
     \end{cases}
\end{align}
The threshold in \eqref{eq:threshold} can be easily obtained as $\hat{\theta}\approx26.56^{\circ}$, which satisfies $\tan{\hat{\theta}}=\frac{1}{2}$. Note that the threshold is biased to use $\mathcal{X}_2$ due to the different instantaneous transmission power levels. From \eqref{eq:threshold}, the lower bound of \eqref{eq:SISO_rerepresent} is obtained as
\begin{align}
    C^{\sf SISO}\geq R^{\sf SISO}\stackrel{(a)}{\geq} \begin{cases}
        2\left(1\!-\!{\sf H}_b\left(Q\!\left(\sqrt{\frac{2}{\sigma^2}({\bar h}_{\rm Re}^2\!+\!{\bar h}_{\rm Im}^2) \sin^2 \theta}\right)\right)\right), \ \ \ \ \ \ \text{if } \ \theta\!>\!26.56^{\circ},\\
         2\left(1\!-\!{\sf H}_b\left(Q\!\left(\sqrt{\frac{2}{\sigma^2}({\bar h}_{\rm Re}^2\!+\!{\bar h}_{\rm Im}^2)(1\!-\!\sin2\theta)}\right)\right)\right), \  \text{if } \ \theta\!<\!26.56^{\circ},
    \end{cases}
\end{align}
where (a) is because ${\sf{H}}_b(Q(\sqrt{x}))$ is a decreasing function with respective to $x$. Compared to the capacity with infinite-precision DACs in \eqref{eq:CSIT_SISO}, the power loss factor is at most
\begin{align}
P_{\rm loss}=\max\left\{\sin^2\theta,1-\sin2\theta\right\}\leq 7 \text{dB}.
\label{eq:powerloss}
\end{align}
In addition, if we assume that the channel phase is uniformly distributed, i.e., Rayleigh fading channel environments, the power loss due to the use of one-bit DACs in an ergodic sense is
\begin{align}
   \mathbb{E}_{\bf h}\left[P_{\rm loss}\right]=\frac{4}{\pi}\left\{\int_{0}^{\hat {\theta}}(1-\sin2\theta)d\theta+\int_{\hat{\theta}}^{\frac{\pi}{4}}\sin^2\theta d\theta\right\}\leq 3.2 \text{dB}.
\end{align}
This result reveals that the loss due to the use of one-bit DACs with one-bit feedback is not significant compared to the case of using infinite-bit DACs with the infinite amount of feedback bits.

\vspace{-0.1cm}
\section{Capacity of MISO Channels with One-Bit Transceiver}

In this section, we extend the results in Section III for the MISO channel with one-bit ADCs and DACs.
\vspace{-0.2cm}
\subsection{Capacity of the MISO Channel}
We first present our main result of this section.

\vspace{0.1cm}
{\bf Theorem 2:} For a given channel realization, i.e., ${\bf H}=\mathcal{H}$, the capacity of the MISO channel with one-bit ADCs and DACs is
\begin{align}
	C^{\sf MISO}= 2-\sum_{u=1}^{2M}\sum_{k=1}^{K_u} {p^{\star}_{u,k}}{\sf H}_b^{\mathcal{X}_{u,k}} , \label{eq:MISO_Capacity}
\end{align}	\vspace{-0.1cm}where $p_{u,k}^{\star}$ is the optimal solution of the following linear programming problem:
\begin{align} \label{eq:optimization}
   &\min_{p_{1,1},\ldots, p_{2M,K_{2M}}}~~  \sum_{u=1}^{2M}\sum_{k=1}^{K_u} p_{u,k}{\sf H}_b^{\mathcal{X}_{u,k}} \nonumber\\
 &~~~~~~\textrm{such that}~~ \sum_{u=1}^{2M}u\sum_{k=1}^{K_u} p_{u,k} \leq P_{\rm t}, \nonumber\\
  &~~~~~~~~~~~~~~~~~~~ \sum_{u=1}^{2M}\sum_{k=1}^{K_u} p_{u,k} =1, \nonumber\\
 &~~~~~~~~~~~~~~~~~~~ p_{u,k}\geq 0.
\end{align} \label{eq:MISO_Capacity_1_LP}\vspace{-0.4cm}
	\begin{IEEEproof}
The proof resembles with that of the SISO case. Using the rotationally invariant property of the optimal input distribution shown in Lemma 1, we consider the input distribution, which is uniformly distributed over four symmetric input vectors in $\mathcal{X}_{u,k}$, i.e., 
\begin{align}
	{\rm Pr}\left[{\bf x}={\bf R}^i{\bf x}_{u,k}\right]=\frac{p_{u,k}}{4}, \ \forall i \in \{0,1,2,3\},
\end{align}
for $u \in \{1,2,\ldots,2M\}.$
Recall that the input-output relationship in \eqref{eq:real_inoutput} for the case of the MISO channel is
\begin{align}
\begin{bmatrix}
	y_1 \\
	y_2
\end{bmatrix}=
	\begin{bmatrix}
	{\sf sign}\left({\bf h}_{1}^{\top}{\bf x}+z_{1}\right) \\
	{\sf sign}\left({\bf h}_{2}^{\top}{\bf x}+z_{2}\right)
\end{bmatrix}.
\end{align}
Then, the mutual information between the input ${\bf x}$ and the output ${\bf y}$ when the channel is given by ${\bf H}=\mathcal{H}$ is
\begin{align}
	{\sf I}({\bf x};{\bf y}|{\bf H}=\mathcal{H})={\sf H}({\bf y}|{\bf H}=\mathcal{H}) - {\sf H}({\bf y}|{\bf x},{\bf H}=\mathcal{H}) \label{eq:MI_MISO}.
\end{align}
To compute ${\sf H}({\bf y}|{\bf H}=\mathcal{H})$ in \eqref{eq:MI_MISO}, we first compute
\begin{align}
	{\rm Pr}\left[{\bf y}=[1,1]^{\top}|{\bf H}=\mathcal{H}\right]
	&=\sum_{u=1}^{2M}\sum_{k=1}^{K_u}\sum_{i=0}^{3} {\rm Pr}\left[{\bf y}=[1,1]^{\top}|{\bf x}={\bf R}^i{\bf x}_{u,k}, {\bf H}=\mathcal{H}\right]{\rm Pr}\left[{\bf x}={\bf R}^i{\bf x}_{u,k}\right]  \nonumber \\
	&\stackrel{(a)}{=}\sum_{u=1}^{2M}\sum_{k=1}^{K_u} \sum_{i=0}^{3}\left(\prod_{n=1}^2{\rm Pr}\left[y_n=1|{\bf x}={\bf R}^i{\bf x}_{u,k},{\bf h}_n^{\top}\right]\right){\rm Pr}\left[{\bf x}={\bf R}^i{\bf x}_{u,k}\right]\nonumber\\
	&\stackrel{(b)}{=}\sum_{u=1}^{2M}\sum_{k=1}^{K_u} \frac{p_{u,k}}{4}\sum_{i=0}^{3}\prod_{n=1}^2{\rm Pr}\left[y_n=1|{\bf x}={\bf R}^i{\bf x}_{u,k},{\bf h}_n^{\top}\right],\label{eq:entropy_y_MISO}
	\end{align}
where (a) is because the independence of $z_1$ and $z_2$ and (b) follows from Lemma 1. By expansion we rewrite
\begin{align}
    &\sum_{i=0}^{3}\prod_{n=1}^2{\rm Pr}\left[y_n=1|{\bf x}={\bf R}^i{\bf x}_{u,k},{\bf h}_n^{\top}\right]=1,
\end{align}
using the similar approach in \eqref{eq:expand}. Therefore, we obtain
\begin{align}
	{\rm Pr}\left[{\bf y}=[1,1]^{\top} |{\bf H}=\mathcal{H}\right]=\sum_{u=1}^{2M}\sum_{k=1}^{K_u} \frac{p_{u,k}}{4}=\frac{1}{4}.
	\end{align}
By symmetry, the channel outputs are uniformly distributed regardless of the channel values. Consequently, the channel output entropy becomes
\begin{align}
		{\sf H}({\bf y}|{\bf H}=\mathcal{H})=2.
\end{align}
Now, we calculate the conditional entropy given ${\bf x}$ as a form of the weighted sum of binary entropy functions as
\begin{align}
    {\sf H}\left(y_n|{\bf x},{\bf H}=\mathcal{H}\right)&=\sum_{u=1}^{2M}\sum_{k=1}^{K_u} \frac{p_{u,k}}{4}\sum_{i=0}^{3}\left\{{\sf H}\left(\!y_1|{\bf x}={\bf R}^i{\bf x}_{u,k},{\bf h}_{1}^{\top}\right)+{\sf H}\left(y_2|{\bf x}={\bf R}^i{\bf x}_{u,k},{\bf h}_{2}^{\top}\right)\right\}\nonumber\\&{=}\sum_{u=1}^{2M}\sum_{k=1}^{K_u}p_{u,k}\left\{{\sf H}\left(y_1|{\bf x}={\bf x}_{u,k},{\bf h}_{1}^{\top}\right)+{\sf H}\left(y_2|{\bf x}={\bf x}_{u,k},{\bf h}_{2}^{\top}\right)\right\}\nonumber\\ &{=}\sum_{u=1}^{2M}\sum_{k=1}^{K_u}p_{u,k}\sum_{n=1}^{2}{\sf H}_b\left(Q\left(\sqrt{\frac{2}{\sigma^2}}{\bf h}_n^{\top}{\bf x}_{u,k}\right)\right).
\end{align}
As a result, the mutual information of the MISO channel with one-bit transceivers is 
\begin{align}
	{\sf I}({\bf x};{\bf y}|{\bf H}=\mathcal{H})&={\sf H}({\bf y}|{\bf H}=\mathcal{H}) - \sum_{n=1}^2 {\sf H}\left(y_n|{\bf x},{\bf H}=\mathcal{H}\right)  \nonumber\\
	&=2-\sum_{u=1}^{2M}\!\sum_{k=1}^{K_u}p_{u,k}\sum_{n=1}^{2}{\sf H}_b\left(Q\left(\sqrt{\frac{2}{\sigma^2}}{\bf h}_n^{\top}{\bf x}_{u,k}\right)\right)=2-\sum_{u=1}^{2M}\sum_{k=1}^{K_u} {p_{u,k}}{\sf H}_b^{\mathcal{X}_{u,k}}  .\label{eq:MI_MISO2} 
\end{align}
By invoking the optimal distribution from \eqref{eq:optimization} into \eqref{eq:MI_MISO2}, we obtain the capacity expression in Theorem 2. This completes the proof.   \end{IEEEproof}
\vspace{-0.2cm}
\subsection{Implication}
The capacity expression in \eqref{eq:MISO_Capacity} is unwieldy to attain a clear intuition, because the optimal solution of $p_{u,k}^{\star}$ depends on the average power constraint $P_{\rm t}$. To avoid this, we focus on the case when the average transmission power is sufficient to use the signals sets with the instantaneous power of $2M$, i.e., $P_{\rm t}=2M$. In this case, the capacity expression is simplified as the following corollary.

{\bf Corollary 4:} When $P_{\rm t}= 2M$, the capacity of the MISO channel with one-bit transceivers is 
\begin{align}
	C^{\sf MISO}&=  2-\min_{u,k}\left\{{\sf H}_b^{\mathcal{X}_{u,k}}\right\}.
	 \label{eq:capacity_MISO_CF}
\end{align}
\begin{IEEEproof}
When $P_{\rm t}= 2M$, any signal SLM subset $\mathcal{X}_{u,k}$ can be harnessed, i.e., the time-sharing technique is not necessarily needed to satisfy the average transmit power constraint. Consequently, the optimal transmission strategy is to use four rotationally invariant SLM signal points, i.e., $\mathcal{X}_{u,k}$ that minimizes ${\sf H}_b^{\mathcal{X}_{u,k}}$ depending on the channel and the SNR.
\end{IEEEproof}

		{\bf Corollary 5:} When $P_{\rm t}=2M$, $\log_2\left(\frac{9^M-1}{4}\right)$ feedback bits for CSIT are sufficient to achieve the capacity of the MISO channel with one-bit transceivers.
	\begin{IEEEproof}
	The capacity is achievable by sending back the index of the subset $\mathcal{X}_{u,k}$ that yields the minimum value of the average binary entropy functions ${\sf H}_b^{\mathcal{X}_{u,k}}$ from a receiver. Since there exists $\sum_{u=1}^{2M}{2M \choose u}2^{u-2}=\frac{9^M-1}{4}$ number of subsets, $\log_2\left(\frac{9^M-1}{4}\right)$ feedback bits are enough to achieve the capacity in \eqref{eq:capacity_MISO_CF}.
	\end{IEEEproof}
	
{\bf Remark 3 (Capacity loss by the use of one-bit DACs):} 
When infinite-resolution DACs are employed, the capacity is achieved by the QPSK modulation with MRT precoding as shown in \cite{Mo2015_TSP}. In this case, the MISO channel is equivalent to a SISO channel with the channel gain of $\|{\bf h}_1\|_2$. Therefore, the capacity is obtained as
\begin{align}
    	C_{\sf ADCs}^{\sf MISO}=2\left(1-{\sf H}_b\left(Q\!\left(\sqrt{\frac{2}{\sigma^2}\|{\bf h}_1\|^2_2}\!\right)\right)\right). \label{eq:CSIT_MISO}
\end{align}
With \eqref{eq:CSIT_MISO}, the capacity loss due to the use of DACs is
\begin{align}
    \Delta C_{\sf DACs}^{\sf MISO}=\min_{u,k}\left\{  {\sf H}_b^{\mathcal{X}_{u,k}}\right\} -2{\sf H}_b\left(Q\left(\sqrt{\frac{2}{\sigma^2}\|{\bf h}_1\|^2_2}\right)\right).
\end{align}
It is notable that the capacity in \eqref{eq:CSIT_MISO} can only be achievable when infinite amount of CSIT feedback bits are available. Whereas, when using one-bit transceivers, the capacity is achievable with $\log_2\left(\frac{9^M-1}{4}\right)$ feedback bits.

{\bf Remark 4  (Generalization to the multi-bit DACs):} For the case in which multi-bit DACs are employed at the transmitter,  it is possible to generate the channel input set using the SLM method. For example, when two-bits DACs are used, possible channel input set is $\mathcal{X}=\{-2,-1,0,1,2\}^{2M}/\{{\bf 0}\}$, where $|\mathcal{X}|=\sum_{i=0}^{2M}{2M \choose i}5^i-1=5^{2M}-1$. In this case, there exists $\frac{25^M-1}{4}$ number of subsets, i.e., $\mathcal{X}_{u,k}$, each with four SLM signal points that have the same antenna activation pattern. By deriving the optimal time-sharing transmission scheme among the $\frac{25^M-1}{4}$ number of possible subsets under the average power constraint, one can find the capacity expression for the MISO channel with multi-bit DACs and one-bit ADCs.

\vspace{-0.1cm}
\section{Channel Training and Feedback}	
 In this section, we propose a practical capacity-achieving downlink transmission technique including channel training and feedback methods for MISO channels with one-bit transceiver. For simplicity, we focus on the case of $P_{\rm t}=2M$, which enables us to use the simple capacity expression in Corollary 4.

 The key idea of our proposed strategy is to empirically learn the entropy functions for each SLM subsets, i.e., ${\sf H}_b^{\mathcal{X}_{u,k}}$ for $u,k$, by repeatedly sending the input vectors in $\mathcal{X}_{u,k}$ as a training sequence. This idea extends an implicit channel-learning method developed in our prior work \cite{Jeon2017_SL,Jeon2018_SL,Jeon_So_Lee_2018}. This strategy allows the receiver to estimate ${\sf H}_b^{\mathcal{X}_{u,k}}$ directly instead of estimating the downlink channel itself ${\bf \bar h}^{\top}\in \mathbb{C}^{1\times M}$. In \cite{Jeon2017_SL,Jeon2018_SL}, this implicit channel-learning method is shown to be effective compared to the direct channel estimation, because the accuracy of the direct channel estimation method is poor when using one-bit ADCs at the receiver. Then, using the estimate of  ${\sf H}_b^{\mathcal{X}_{u,k}}$, the receiver determines the best SLM subset, which achieves the capacity. Once the optimal index of the SLM subset is found, the receiver sends it back to the transmitter via a finite-rate feedback link.  To accomplish the proposed strategy, we present two channel-training-and-feedback methods, referred to as \textit{full training} and \textit{dominant-set training}.




\vspace{-0.1cm}
\subsection{Full Training}
In the full training method, the transmitter first sends $L$ repetitions of all constellation vectors $\{{\bf x}_{u,k}\}_{u,k}$ for $k\in\{1,\ldots,K_u\}$ and $u\in\{1,\ldots,2M\}$ as a training sequence. In this case, the training length becomes $\frac{9^M-1}{4}L$. This training sequence allows the receiver to obtain $L$ received vectors, namely, $\{{\bf y}_{u,k}[\ell]\}_{\ell=1}^{L}$, associating with each channel input vector ${\bf x}_{u,k}$. Using these $L$ observations, the receiver empirically computes the likelihood function of $y_n$ for given ${\bf x}_{u,k}$ as
\begin{align} 
	{\rm Pr}[y_n=1|{\bf x}={\bf x}_{u,k}, {\bf h}_n^{\top}]= {\rm Pr}[ z_n > -{\bf h}_n^{\top}{\bf x}_k] 
	&=\!1\!-\!Q\left(\!\sqrt{\frac{2}{\sigma^2}}{\bf h}_n^{\top}{\bf x}_{u,k}\!\right)\simeq \frac{1}{L}\sum_{\ell=1}^{L}{\bf 1}_{\{y_{u,k,n}[\ell]=1\}},
	\label{eq:emp_pro}
\end{align}	
where ${y}_{u,k,n}[\ell]$ is the $n$-th element of ${\bf y}_{u,k}[\ell]$ and ${\bf 1}\{\cdot\}$ is an indicator function that yeilds 1 if an event is true, and zero otherwise. From \eqref{eq:emp_pro}, the receiver also computes the empirical entropy function for the input set $\mathcal{X}_{u,k}$ as
\begin{align}
	{\sf H}_b^{\mathcal{X}_{u,k}} = \sum_{n=1}^{2}{\sf H}_b\left(Q\left(\sqrt{\frac{2}{\sigma^2}}{\bf h}_n^{\top}{\bf x}_{u,k}\right)\right)
	\!\overset{(a)}{\simeq} \sum_{n=1}^{2}{\sf H}_b\left(\frac{1}{L}\sum_{\ell=1}^{L}{\bf 1}\big\{{y}_{u,k,n}[\ell]=1\big\}\right), \label{eq:emp_ent} 
\end{align}
where (a) follows from ${\sf H}_b(x)={\sf H}_b(1-x)$. Since we assume that the transmit power is sufficient (i.e., $P_{\rm t}=2M$) as in Corollary 4, the index of the capacity-achieving input set is determined as
\begin{align}
	(u^\star, k^\star) 
	&= \argmin_{u,k} \sum_{n=1}^{2}{\sf H}_b\left(\frac{1}{L}\sum_{\ell=1}^{L}{\bf 1}\big\{{y}_{u,k,n}[\ell]=1\big\}\right). \label{eq:emp_best} 
\end{align}
Finally, the receiver sends back the best index $(u^\star, k^\star)$ to the transmitter via a finite-rate feedback link, which requires the rate of $\log_2{\sum_{u=1}^{2M}{K_u}}\approx 3.17M-2$ feedback bits/sec. One feature of the full training method is that when the number of training repetitions $L$ is sufficiently large, the best input set determined from \eqref{eq:emp_best} is indeed the capacity-achieving input set.

\vspace{-0.1cm}
\subsection{Dominant-Set Training}
One drawback of the full training method is that its training overhead exponentially increases with the number of transmit antennas, i.e., $\frac{9^M-1}{4}L$. This overwhelming overhead drastically reduces the throughput of wireless systems, particularly when the number of transmit chains is large. To resolve this problem, we also present a dominant-set training method, which requires a less training overhead than the full training method.

The idea of the proposed dominant-set-training method is to empirically estimate ${\sf H}_b^{\mathcal{X}_{2M,k}}$ for the channel input sets, which have the maximum instantaneous transmission power, $\{\mathcal{X}_{2M,k}\}_k \subset \{-1,+1\}^{2M}$. Notice that the channel input sets $\mathcal{X}_{2M,k}$ have a more chance to be selected as the capacity-achieving input set in the low SNR regime, because they use a higher instantaneous power than the other input sets. Using this fact, in the dominant-set training method, the transmitter sends $L$ repetitions of the channel input vectors $\{{\bf x}_{2M,k}\}_{k}$ only. Due to the reduced number of the possible channel inputs, the training overhead is considerably reduced to $K_{2M}L=4^{M-1}L$, which is significantly less than that of the full-training method. From the reduced training sequence, the receiver computes the empirical entropy functions for $\{\mathcal{X}_{2M,k}\}_k$ and then determines the index of the best input set as
\begin{align}
	k^\star 
	&=  \argmin_{k} \sum_{n=1}^{2}{\sf H}_b\left(\frac{1}{L}\sum_{\ell=1}^{L}{\bf 1}\big\{{y}_{2M,k,n}[\ell]=1\big\}\right). \label{eq:emp_best_dom} 
\end{align}
Finally, the receiver sends back the best index $k^\star$ to the transmitter via a feedback link with a finite rate of $\log_2{{K_{2M}}}=2M-2$ bits/sec/Hz. One feature of the dominant-training method is that it is able to diminish both the training overhead and the number of feedback bits compared to the full training method at the cost of the performance degradation.

\section{Simulation Results}
\begin{figure}
\centerline{\includegraphics[width=9.0cm]{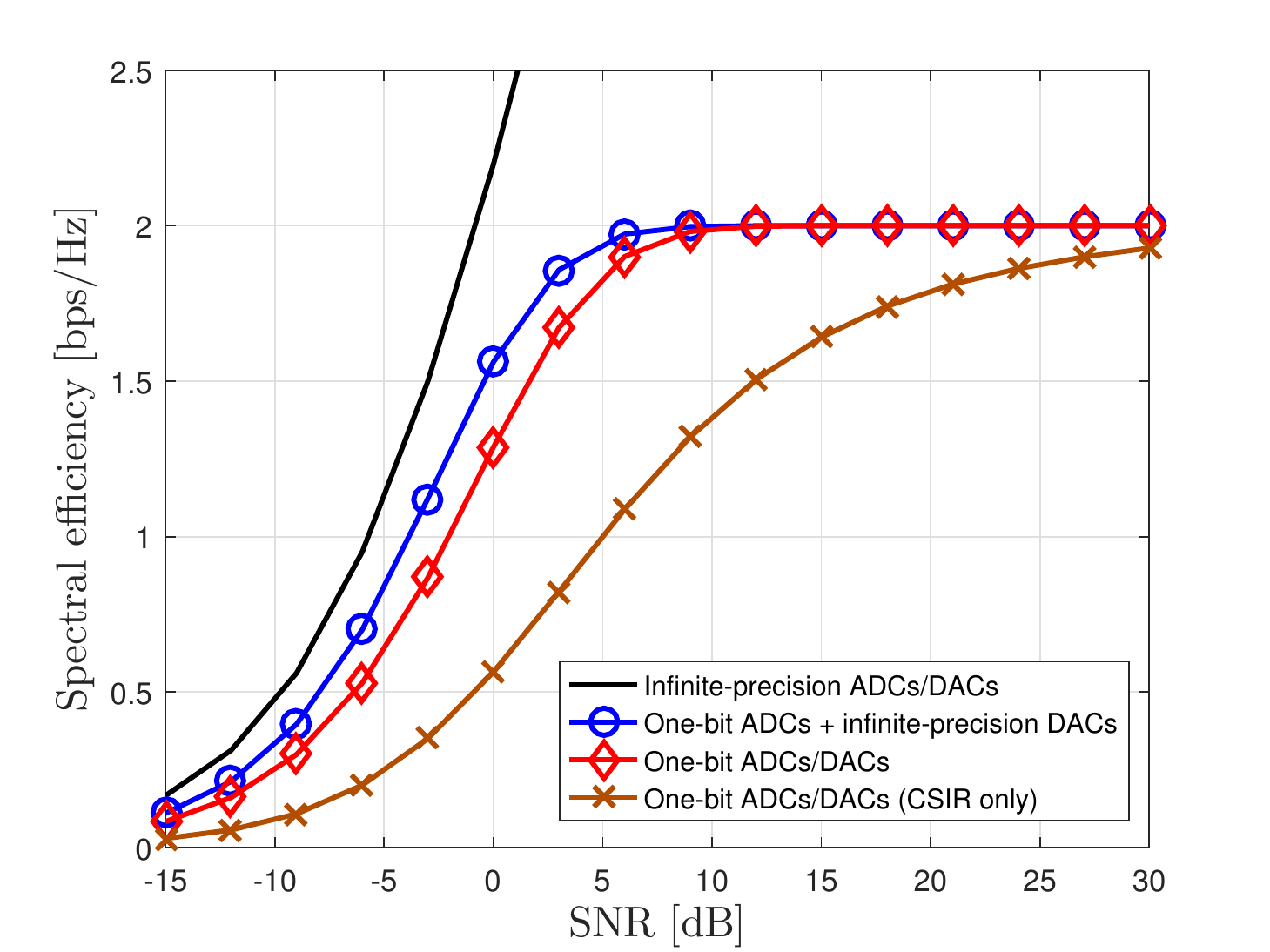}}\vspace{-0.4cm}
\caption{Ergodic capacity comparison of the MISO fading channel when $M=4$ with the different types of ADCs and DACs precision.}
\label{model}\vspace{-0.5cm}
\end{figure}

In this section, using simulations, we evaluate the ergodic capacity of the MISO channel with one-bit transceivers. In our simulation,  we define the ${\rm SNR}=\frac{2M}{\sigma^2}$ by setting $P_{\rm t}=2M$. To evaluate the ergodic capacity, each element of the channel is assumed to be distributed as a circularly-symmetric complex Gaussian random variable with zero-mean and unit variance, i.e., $\mathcal{CN}(0,1)$.

{\bf Effects of one-bit ADCs and DACs:} We first evaluate the ergodic capacity of the Rayleigh MISO channels. To provide a complete picture on how the ADCs and DACs affect the capacity of the MISO channel, we consider the four possible combinations: 1) infinite-precision ADCs and DACs, 2) one-bit ADCs and infinite-precision DACs, 3) one-bit ADCs and DACs, 4) one-bit ADCs and DACs with CSIR only. As can be seen in Fig. 5, the spectral efficiency when using both infinite-precision ADCs and DACs increases with the SNR. Whereas, the spectral efficiencies for the other cases are limited by 2 bits/sec/Hz in the high SNR because of a finite number of the channel input or output values imposed by one-bit ADCs or DACs. One interesting observation is that when the perfect knowledge of CSIT is given, the capacity loss by the use of one-bit DACs is about 2dB in the mid SNR region, compared to the case when the infinite-precision DACs are employed at the transmitter. In addition, it is observed that, for the wireless system using one-bit transceivers, exploiting CSIT is able to improve the spectral efficiency significantly compared to the case where CSIR is only available. 
 
\begin{figure}
    \centerline{\includegraphics[width=9.0cm]{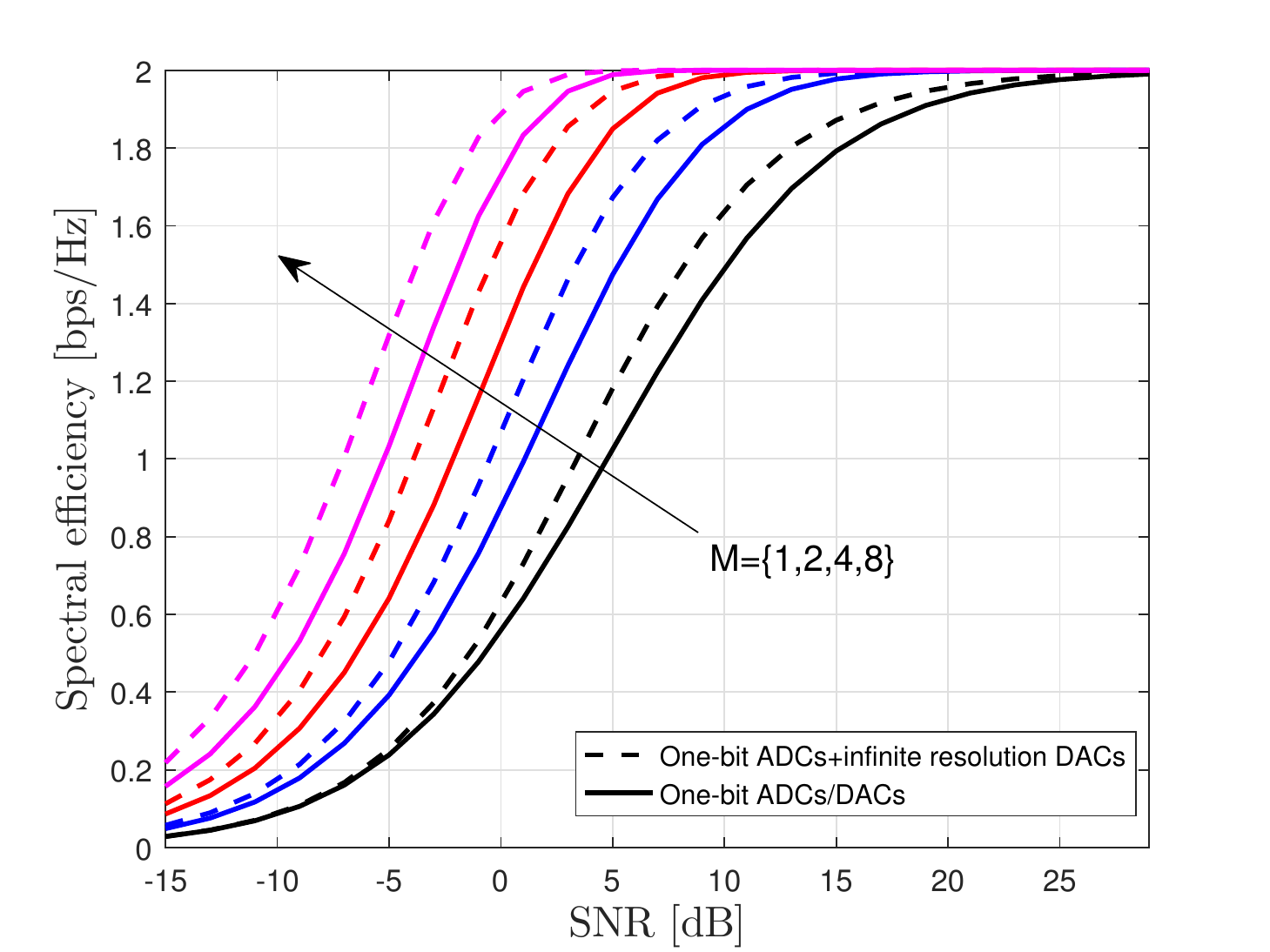}}\vspace{-0.4cm}
    \caption{Ergodic capacity comparison of the MISO fading channel when increasing $M$.}
    \label{fig:FullTrain}\vspace{-0.5cm}
\end{figure}

{\bf Effects of the number of transmit antenna chains:} Fig. 6 illustrates how the number of the transmit antennas $M$, each with one-bit DACs changes the ergodic capacity. As $M$ increases, it is shown that the capacity increases considerably, similar to the case when the transmitter uses infinite-precision DACs.  The capacity improvement is possible because of the transmit diversity gain. Since the number of possible channel input sets exponentially increases with $M$, it is highly likely to find a channel input set that is well aligned with the channel direction.


\begin{figure}
    \centerline{\includegraphics[width=9cm]{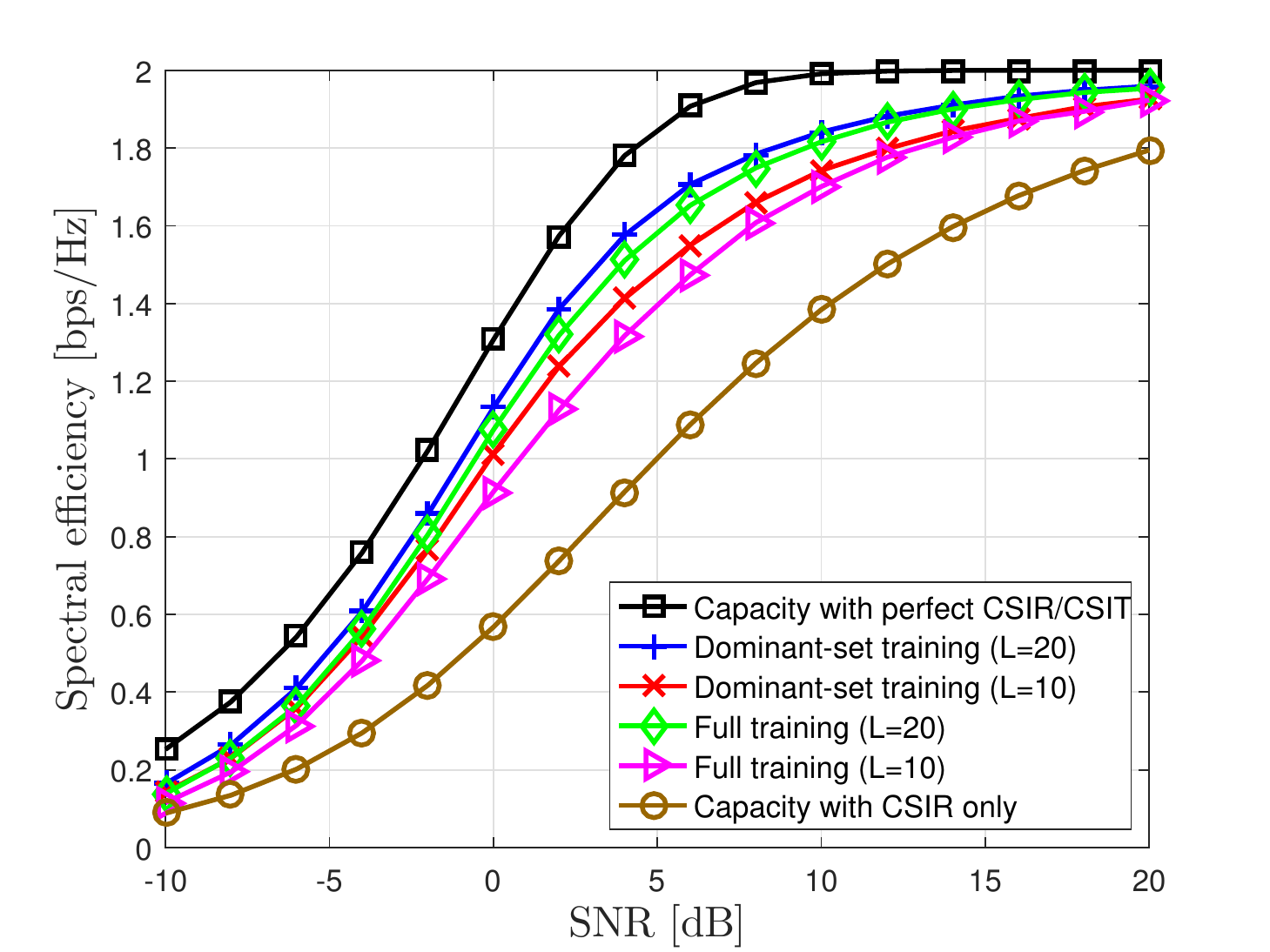}}\vspace{-0.4cm}
    \caption{Ergodic capacity comparison of the MISO fading channel when $M=4$ for different channel training methods.}
    \label{fig:DomTrain}\vspace{-0.5cm}
\end{figure}

{\bf Effects of imperfect CSIR and CSIT:} We also characterize how the imperfect CSIR and CSIT have an effect on the capacity of a MISO Rayleigh fading channel when the proposed methods explained in Section V are employed. The numbers of repetitions per a channel input vector are set to be $L=10$ and $L=20$, respectively. Since there is a total of $1640$ distinct SLM subsets, i.e., $\mathcal{X}_{u,k}$ when $M=4$,  the full training method with $L=10$ and $L=20$ requires the training overheads of $16400$ and $32800$, respectively. The corresponding feedback amount of the full-training method becomes $10.68$ bits, which is sent to the transmitter via an error-free feedback link. Notice that the CSIT obtained by the full-training method is imperfect, even if we use the error-free feedback link. This is because the optimal index of the channel input subset can be chosen with errors by the channel training procedure.  Whereas, the dominant-set training method is able to drastically diminish the overheads for the channel training and feedback. Specifically, this method with $L=10$ and $L=20$ needs the training overheads of $640$ and $1280$, respectively, and $6$-bit feedback is needed. As can be seen in Fig.~\ref{fig:DomTrain}, the proposed dominant-set training method achieves a close performance to the capacity even with a reasonable number of training overheads. One interesting observation is that the dominant-set training method even outperforms the full training method. This count-intuitive result is because the full training method considers all possible input sets, and the most of them are unlikely to be chosen as the capacity-achieving input set, if the average transmit power is set to be $P_{\rm t}=2M$. This effect is shown to be more critical in a low SNR regime, in which the probability of the incorrect input selection is high. From the numerical results, it is observed that the dominant-set training method not only reduces the training and feedback overheads, but also can improve the achievable spectral efficiency compared to the full training method in a practical scenario.

\vspace{-0.1cm}
\section{Conclusion}

The key features of future wireless systems are the use of a large signal bandwidth and massive antennas to support very high data transmission. Exploiting such wide-band signal and massive antenna arrays in the design of wireless systems may cause extremely high power consumption and device costs. This paper focused on the simplest yet effective solution by using one-bit DACs and ADCs at the transmit and receive chains. Most notably, information-theoretical limits of the MISO channel using one-bit transceiver  were characterized in closed forms when both perfect CSIT and CSIR are available. The key finding was that the capacity-achieving scheme is a time sharing technique among multiple SLM subsets. With the found capacity expressions, it has been also revealed that a finite-rate of CSIT feedback is sufficient. Aside from the capacity characterization, a practical transmission strategy was also presented using a supervised-learning approach when both imperfect CSIR and CSIT are given. Using simulations, it was demonstrated that the proposed strategy achieves a close performance to the capacity with a reasonable number of training and feedback overheads.  

An important direction for future work would be to characterize the fundamental limit of the MIMO channel using one-bit transceivers. For the MIMO channel, from the result of Lemma 1, we conjecture that the capacity-achieving transmission method is to use multiple SLM subsets, in which the elements of each subset are uniformly distributed. Another important extension is to design multi-user precoding methods for downlink multi-user MIMO systems with one-bit DACs \cite{Usman,Swindlehurst,Yu} by leveraging the SLM method \cite{Choi2018}.


\vspace{-0.1cm}
\appendices
 
 \section{Proof of Lemma 1}
 \vspace{-0.1cm}
 {\bf Lemma 1:} For a fixed MIMO channel ${\bf H}=\mathcal{H}\in \mathbb{R}^{2N\times 2M}$ with one-bit ADCs and DACs, there exists a capacity-achieving input distribution, which is uniform in each $\mathcal{X}_{u,k}$, i.e.,
 \begin{align}
    	{\rm Pr}[{\bf x}={\bf R}^i{\bf x}_{u,k}]=\frac{p_{u,k}}{4}, \ \forall i \in \{0,1,2,3\},
\end{align}
for $u \in \{1,2,\ldots,2M\}$.
 \begin{IEEEproof}
 For any rotation matrix, i.e., ${\bf R}^i$, and any initial input distribution, i.e., ${\bf x}\sim {\rm Pr}_0[{\bf x}]$, we define a distribution for the rotated input by ${\rm Pr}_i[{\bf x}]={\rm Pr}_0[{\bf R}^i{\bf x}]$. We first show that the mutual information between ${\bf x}$ and ${\bf y}$ is preserved under the input rotation. As an example, with the input distribution of ${\rm Pr}_1[{\bf x}]$, we obtain
 \begin{align}
  {\bf y}&={\sf sign}\left( \begin{bmatrix}
  {\bf \bar H}_{\rm Re} &  -{\bf \bar H}_{\rm Im} \\      
 {\bf \bar H}_{\rm Im} &  {\bf \bar H}_{\rm Re} \\
 \end{bmatrix}\!
 \begin{bmatrix}
 {\bf 0}_M & -{\bf I}_M \\
 {\bf I}_M & {\bf 0}_M
 \end{bmatrix}\!
 \begin{bmatrix}
 \bar{\bf x}_{\rm Re} \\
 \bar{\bf x}_{\rm Im}
 \end{bmatrix}\!
 +\!{\bf z}\right)=
 {\sf sign}\left(\begin{bmatrix}
  -{\bf \bar H}_{\rm Im} &  -{\bf \bar H}_{\rm Re} \\      
 {\bf \bar H}_{\rm Re} &  -{\bf \bar H}_{\rm Im} \\
 \end{bmatrix}\!
 \begin{bmatrix}
 \bar{\bf x}_{\rm Re} \\
 \bar{\bf x}_{\rm Im}
 \end{bmatrix}\!
 +\!{\bf z}\right)\nonumber\\
 &=
 {\sf sign}\left(\begin{bmatrix}
		{\bf 0}_{N} & -	{\bf I}_{N} \\
		{\bf I}_{N} & 	{\bf 0}_{N} \\
\end{bmatrix}\begin{bmatrix}
  {\bf \bar H}_{\rm Re} &  -{\bf \bar H}_{\rm Im} \\      
 {\bf \bar H}_{\rm Im} &  {\bf \bar H}_{\rm Re} \\
 \end{bmatrix}\!
 \begin{bmatrix}
 \bar{\bf x}_{\rm Re} \\
 \bar{\bf x}_{\rm Im}
 \end{bmatrix}\!
 +\!{\bf z}\right)=
 {\sf sign}\left(\begin{bmatrix}
		{\bf 0}_{N} & -	{\bf I}_{N} \\
		{\bf I}_{N} & 	{\bf 0}_{N} \\
\end{bmatrix}
\left(
 \begin{bmatrix}
  {\bf \bar H}_{\rm Re} &  -{\bf \bar H}_{\rm Im} \\      
 {\bf \bar H}_{\rm Im} &  {\bf \bar H}_{\rm Re} \\
 \end{bmatrix}\!
 \begin{bmatrix}
 \bar{\bf x}_{\rm Re} \\
 \bar{\bf x}_{\rm Im}
 \end{bmatrix}\!
 +\!\begin{bmatrix}
		{\bf 0}_{N} & {\bf I}_{N} \\
		-{\bf I}_{N} & 	{\bf 0}_{N} \\
\end{bmatrix}\!{\bf z}\right)\right) \nonumber\\
&=
\begin{bmatrix}
		{\bf 0}_{N} & -{\bf I}_{N} \\
		{\bf I}_{N} & 	{\bf 0}_{N} \\
\end{bmatrix}{\sf sign}
\left(
 \begin{bmatrix}
  {\bf \bar H}_{\rm Re} &  -{\bf \bar H}_{\rm Im} \\      
 {\bf \bar H}_{\rm Im} &  {\bf \bar H}_{\rm Re} \\
 \end{bmatrix}\!
 \begin{bmatrix}
 \bar{\bf x}_{\rm Re} \\
 \bar{\bf x}_{\rm Im}
 \end{bmatrix}\!
 +\!\begin{bmatrix}
		{\bf 0}_{N} & {\bf I}_{N} \\
		-{\bf I}_{N} & 	{\bf 0}_{N} \\
\end{bmatrix}\!{\bf z}\right).
\end{align}
 Since the noise is circularly-symmetric, we obtain ${{\sf I}_1\left({\bf x};{\bf y}|{\bf H}=\mathcal{H} \right)}={{\sf I}_0\left({\bf x};{\bf y}|{\bf H}=\mathcal{H} \right)}$, where ${{\sf I}_i\left({\bf x};{\bf y}|{\bf H}=\mathcal{H} \right)}$ denotes the mutual information between the input and the output under ${\rm Pr}_i[{\bf x}]$. In a similar manner, it is possible to show
 \begin{align}
    {{\sf I}_i\left({\bf x};{\bf y}|{\bf H}=\mathcal{H} \right)}={{\sf I}_0\left({\bf x};{\bf y}|{\bf H}=\mathcal{H} \right)}, \label{eq:invariance}
\end{align}
for $i \in \{0,1,2,3\}$. Next, we define a convex combination of ${\rm Pr}_i[{\bf x}]$ as
\begin{align}
    {\rm Pr}[{\bf x}]=\frac{1}{4}\sum_{i=0}^{3}{\rm Pr}_i[{\bf x}]. \label{eq:convexcomb}
\end{align} 
 Since the mutual information is a concave function, by the Jensen's inequality, we obtain an upper bound of ${{\sf I}_0\left({\bf x};{\bf y}|{\bf H}=\mathcal{H} \right)}$ as
 \begin{align}
	{{\sf I}_0\left({\bf x};{\bf y}|{\bf H}=\mathcal{H} \right)}&=\frac{1}{4} \sum_{i=0}^3 {{\sf I}_i\left({\bf x};{\bf y}|{\bf H}=\mathcal{H} \right)}\leq  {{\sf I}\left({\bf x};{\bf y}|{\bf H}=\mathcal{H} \right)}. \label{eq:concave_bound}
\end{align}
  As a result, if ${\rm Pr}_0[{\bf x}]$ is a capacity-achieving distribution, ${\rm Pr}[{\bf x}]$ also achieves the capacity, which is uniformly distributed in ${\mathcal X}_{u,k}$ by the definition in \eqref{eq:convexcomb}. This completes the proof.
 \end{IEEEproof}
 
 {\bf Example 3 :} 
To give a clear understanding on Lemma 1, when $M=1$, we suppose the following input distributions ${\rm Pr}_i[{\bf x}]$:
\begin{align}
    	&\left({\rm Pr}\Big[{\bf x}=\begin{bmatrix}
  1 \\      
 0  \\
 \end{bmatrix}\Big]
 , {\rm Pr}\Big[{\bf x}=\begin{bmatrix}
 0 \\      
 1  \\
 \end{bmatrix}\Big], {\rm Pr}\Big[{\bf x}=\begin{bmatrix}
 -1 \\      
 0  \\
 \end{bmatrix}\Big], {\rm Pr}\Big[{\bf x}=\begin{bmatrix}
 0 \\      
 -1  \\
 \end{bmatrix}\Big]\right)\nonumber\\&\!=\!\underbrace{\left(\frac{1}{8},\frac{1}{4},\frac{1}{4},\frac{3}{8}\right)}_{{\rm Pr}_0[{\bf x}]}, \!\underbrace{\left(\frac{1}{4},\frac{1}{4},\frac{3}{8},\frac{1}{8}\right)}_{{\rm Pr}_1[{\bf x}]}, \! \underbrace{\left(\frac{1}{4},\frac{3}{8},\frac{1}{8},\frac{1}{4}\right)}_{{\rm Pr}_2[{\bf x}]}, \textrm{ and } \underbrace{\left(\frac{3}{8},\frac{1}{8},\frac{1}{4},\frac{1}{4}\right)}_{{\rm Pr}_3[{\bf x}]}. \label{eq:equivalent distributions}
\end{align} 
From the property in \eqref{eq:invariance}, these four input distributions achieve the same mutual information between the input and the output. As in \eqref{eq:convexcomb}, by averaging the four input distributions, we obtain the input distribution ${\rm Pr}[{\bf x}]$ as
\begin{align}
    {\rm Pr}[{\bf x}]=\frac{1}{4}\sum_{i=0}^{3}{\rm Pr}_i[{\bf x}]=\left(\frac{1}{4},\frac{1}{4},\frac{1}{4},\frac{1}{4}\right). \label{eq:uniform input}
\end{align}
Here, the inequality in \eqref{eq:concave_bound} ensures that the input distribution in \eqref{eq:uniform input} achieves the higher mutual information than the input distributions in \eqref{eq:equivalent distributions}. Furthermore, as ${\rm Pr}[{\bf x}]$ is an input distribution that is averaged over a rotation matrix ${\bf R}$, it automatically satisfies the uniform property stated in Lemma 1.

\end{document}